\documentclass[twocolumn,showpacs,preprintnumbers,amsmath,amssymb,nofootinbib]{revtex4}

\usepackage[T2A]{fontenc}
\usepackage[cp1251]{inputenc} 
\usepackage[english]{babel}

\usepackage{graphicx}
\usepackage{hyperref}

\begin{document}
\def\be{\begin{equation}}
\def\ee{\end{equation}}
\def\d{\delta}
\def\t{\tilde}
\def\ds{\displaystyle}

\title[Energy spectra of gamma-rays, electrons and neutrinos]%
{Energy spectra of gamma-rays, electrons and neutrinos produced at
proton-proton interactions in the very high energy regime}

\author{S. R. Kelner}
\altaffiliation{Moscow Institute of Engineering Physics, Kashirskoe
sh. 31, Moscow, 115409 Russia}
\email{skelner@rambler.ru}

\author{F. A. Aharonian}
\altaffiliation{Max-Planck-Institut f\"ur Kernphysik,
Saupfercheckweg 1, D-6917 Heidelberg, Germany}
\email{Felix.Aharonian@mpi-hd.mpg.de}

\author{V.~V.~Bugayov}
\altaffiliation{Max-Planck-Institut f\"ur Kernphysik,
Saupfercheckweg 1, D-6917 Heidelberg, Germany}
\altaffiliation{Department of Physics, Altai State University,
656049 Barnaul, Russia}

\date{\today}

\begin{abstract}
We present new parametrisations of energy spectra of
secondary particles, $\pi$-mesons, gamma-rays, electrons and
neutrinos,  produced in inelastic proton-proton collisions. The
simple analytical approximations based on simulations of
proton-proton interactions  using the public available 
SIBYLL  code,  provide very good
accuracy for energy distributions of  secondary products 
in the energy range  above 100 GeV. Generally, the
recommended analytical formulae deviate from the simulated
distributions within a few percent over a large range of $x=E_i/E_p$
- the fraction of energy of the incident proton transferred to the
secondaries. Finally, we describe  an approximate procedure of
continuation of calculations towards low energies, down to the
threshold of $\pi$-meson production.
\end{abstract}


\maketitle

\section{\label{Intro}Introduction}

Any reliable interpretation of an astronomical observation
requires not only high quality experimental information
concerning the spectral, temporal and spatial properties
of radiation, but also unambiguous identification
of  the relevant radiation processes.  In this regard,  the good
knowledge of characteristics of radiation mechanisms is
a principal issue in astrophysics, in particular  in gamma-ray astronomy 
where we often face a problem when the same observation
can be equally well explained by two or more
radiation processes.

Fortunately, all basic radiation processes relevant to high
energy gamma-ray astronomy can be comprehensively studied
using the methods and tools of experimental and theoretical physics.
This concerns, in particular, one of the most principal gamma-ray
production mechanisms  in high-energy astrophysics --
inelastic proton-proton interactions with subsequent decay
of the secondary $\pi^0$ and $\eta$-mesons into gamma-rays. 
The decays of accompanying $\pi^\pm$-mesons and some other 
(less important) short-lived secondaries result in  production
of high energy neutrinos. This  establishes  a deep link between the 
high energy gamma-ray and neutrino astronomies. Finally, the secondary electrons 
(positrons) from p-p interactions may compete with directly accelerated
electrons and thus significantly contribute to the nonthermal electromagnetic
radiation of gamma-ray sources from radio to hard X-rays.

The principal role of this process in high energy astrophysics was 
recognized long ago by the pioneers of the field
(see e.g. \cite{Morrison,GinzSyr,Hayakawa}), in particular  
in the context of their
applications in gamma-ray (e.g. \cite{Stecker_book}) and
neutrino (e.g. \cite{BerVol}) astronomies.
The first reliable experimental results obtained with the SAS-2 and
COS-B gamma-ray satellite missions, initiated new, more detailed
studies of $\pi^0$-production in $p-p$
interactions \cite{StepBad1981,Dermer86},
in particular for the interpretation of the diffuse galactic
gamma-ray background emission. The prospects of
neutrino astronomy initiated similar calculations for
high energy neutrinos and their links 
to gamma-ray astronomy  \cite{Berez_Book,Gaisser,Berez91,Vissani}.

The observations of diffuse gamma-ray emission from different
parts of the galactic disk by EGRET revealed a noticeable excess
of flux at energies of several GeV. Although this excess can be
naturally explained by assuming somewhat harder proton spectrum
(compared to the locally measured flux of cosmic rays) \cite{Mori,AhAt00,StrongMosk00},
this result recently initiated a new study of the process \cite{Kamae05}
based on the approach of separation of the diffractive and non-diffractive
channels of interactions as well as incorporating violation
of the Feynman scaling.

It is remarkable that while precise calculations of gamma-ray spectra
require quite heavy integrations over differential cross-sections
measured at laboratory experiments, the emissivity
of  gamma-rays for an arbitrary broad and smooth, e.g. power-law,
energy distribution of protons can be obtained, with a quite
reasonable accuracy, within a simple formalism (see e.g.  Ref.\cite{AhAt00})
based on the assumption of a constant fraction $\kappa$
of energy of the incident proton released in the secondary gamma-rays
(see \cite{Torres} for comparison of different approaches). 
This approach relies on  the energy-dependent total inelastic 
cross-section of pp interactions, $\sigma_{\rm pp}^{\rm inel}(E)$, 
and assumes  a fixed value of the parameter $\kappa \approx 0.17$,
which  provides the best agreement with the accurate numerical calculations
over a wide  energy range of gamma-rays.

On the other hand, in the case of sharp spectral features  like pileups or cutoffs 
in the proton energy distribution, one has to perform accurate numerical calculations
based on simulations of inclusive cross-sections of production of secondary particles.
Presently three well developed  codes of simulations of \textit{p-p} interactions 
are public available  --  Pythia \cite{Pythia_model}, SIBYLL \cite{SIBYLL_model}, QGSJET \cite{QGSJET_model}. 
The last two as well as some other models are combined in the more general
CORSIKA code \cite{CORSIKA} designed for simulations
of interactions of cosmic rays with the Earth atmosphere.  These codes are based 
on  phenomenological models of  \textit{p-p} interactions incorporated with 
comprehensive  experimental data obtained at particle accelerators.  

While these codes can be directly used for calculations of gamma-ray spectra
for any distribution of primary protons,  it is quite useful to have simple
analytical parameterizations which not only significantly reduce  the calculation
time, but also allow better understanding of characteristics of
secondary electrons, especially for the distributions
of parent protons with distinct spectral features. 
This concerns, for example, such an important question as 
the extraction of the shape of the proton energy distribution 
in the cutoff region based on the analysis of  
the observed gamma-ray spectrum.  Indeed, while the 
delta-functional approach implies, by definition, similar spectral shapes
of gamma-rays and protons (shifted in the energy scale  by a factor of 
$\kappa$), in reality the energy spectrum of highest energy 
gamma-rays appears, as we show below,  smoother than the distribution of
protons in the corresponding (cutoff) region. Thus,  the use of simple approximations
not always can be justified, and, in fact,  they may  cause  
misleading astrophysical  conclusions  about the energy spectra of accelerated protons.

Two different parameterizations of inclusive cross-sections for
pion-production in proton-proton interactions has been published by
Stephens  and Badwar \cite{StepBad1981} and Blattnig et al \cite{Blattnig}.
However,  recently  it was recognized  \cite{Torres} that   both
parameterizations do not  describe correctly  the gamma-ray spectra
in the high energy regime. 
The parameterization of Stephens and Badwar
underpredicts the yield of high energy pions.  The parameterization by 
Blattnig \textit{et al.} is valid for  energies below 50 GeV;  
above this energy  it significantly
overpredicts the pions production. 

In this paper we present new parameterizations for high energy spectra of
gamma-rays, electrons and neutrinos based on the simulations of proton-proton
interactions  using  the SIBYLL code \cite{SIBYLL_model}, and partly  
(only for distributions of $\pi$-mesons) the QGSJET \cite{QGSJET_model} code. 
We provide simple analytical approximations  for the energy spectra of secondaries
with  accuracies  generally better than several percent.

\section{\label{Inclusive} Inclusive spectra of pions}
\begin{figure*}[t]
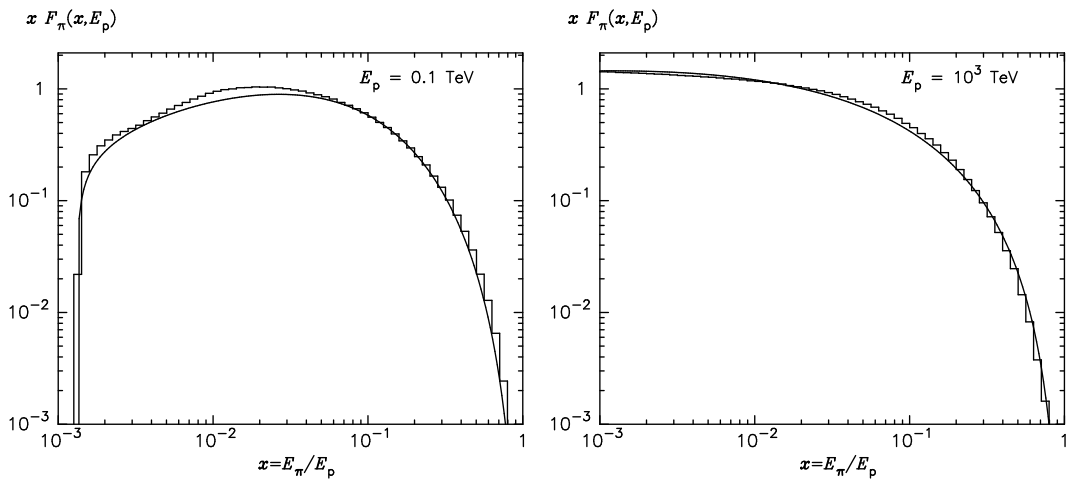

\centering{%
\mbox{\includegraphics[width=0.35\textwidth,angle=-90,clip=]{1a.eps}\quad
\includegraphics[width=0.35\textwidth,angle=-90,clip=]{1b.eps}}
\caption{\small Energy spectra of neutral pions from the
numerical simulations of the QGSJET code (histograms) and from
the presentation given by Eq.(\ref{fitQG2}) for two energies of
primary protons: (a) 0.1~TeV and (b) $10^3$~TeV.}
\label{JETp0_1TeV}
}
\end{figure*}

The energy spectra  of secondary products of  \textit{p-p} interactions 
$F_\pi(x,\,E_p)$ are  expressed  through the total and inclusive
cross-sections, 
\be
\label{inclusive}
 F_\pi(x,\,E_p)=\frac{E_p}{\sigma_{\rm inel}} \int
\frac{d^3 \sigma}{dp^3} \ d^2 p_t^{} \ ,
\ee
where $x=E_\pi/E_p$ is the ratio of the energy of incident proton $E_p$
transferred to the secondary $\pi$-meson.
It is convenient to present $F_\pi(x,\,E_p)$ in the form
\be
\label{fit1}
F_\pi(x,\,E_p)=\frac{d}{dx}\,\Phi(x,\,E_p)\, .
\ee
By definition
\be \label{dNpi} dN_\pi\equiv
F_\pi(x,\,E_p)\,dx=F_\pi(x,\,E_p)\,dE_\pi/E_p
\ee
is the number of
neutral $\pi$-mesons per one $p-p$ interaction in the energy interval
$(E_\pi\,,\, E_\pi+dE_\pi)$.
The presentation in the form of
Eq.(\ref{fit1}) is convenient because it allows to estimate easily
the multiplicity of $\pi$-mesons. The function $\Phi(x,\,E_p)$
obviously can be presented in different forms. For the simulated
distributions obtained with the QGSJET model (see
Fig.~\ref{JETp0_1TeV})  the function 
\be
\label{fitQG1}
\Phi_{\rm QGSJET}^{}=-B_\pi\left(\frac{1-x^\alpha}{(1+r\,x^\alpha)^3}\right)^4\,,
\ee
provides a rather  good fit  with  
the parameters $B_\pi$, $\alpha$ and $r$
as weak functions of $E_p$. Note that
\be
\int\limits_0^1\!F_\pi(x,\,E_p)\,dx=B_\pi\,.
\ee

Also, at the threshold, $x=m_\pi/E_p$,
the function $F_\pi$ should approach zero.
In order to take this effect into account, we introduce
an additional term $(1-m_\pi/(x\,E_p))^{1/2}$ and require that
$F_\pi=0$ at $x<m_\pi/E_p$. Then we obtain
\begin{eqnarray}
F_\pi(x,\,E_p)=4\,\alpha B_\pi
x^{\alpha-1}\left(\frac{1-x^\alpha}{(1+r\,x^\alpha)^3}\right)^4\times\nonumber\\
\left(\frac1{1-x^\alpha}+\frac{3\,r}{1+rx^\alpha}
\right)\left(1-\frac{m_\pi}{x\,E_p}\right)^{1/2}.
\label{fitQG2}
\end{eqnarray}
The spectra of charged pions are described by the same equation.

The parameters $B_\pi$, $\alpha$, and $k$ were obtained
via the best least squares fits to the simulated events (histogram in
Fig.~\ref{JETp0_1TeV}). For the energy interval of
incident protons $0.1 - 10^3$ TeV this procedure gives
\be
B_\pi=5.58+0.78\,L+0.10\,L^2\,,
\ee
\be
r=\frac{3.1}{B_\pi^{3/2}}\,,\quad \alpha=\frac{0.89}{B_\pi^{1/2}
\left(1-e^{-0.33\,B_\pi}\right)}\,,
\ee
where $L=\ln(E_p/1\,{\rm TeV})$.

For $x\agt10^{-3}$  (provided that $E_\pi \geq 10 \ \rm GeV$) 
this analytical approximation deviates from simulations
less than 10 percent for the entire energy range of
incident protons from 0.1 to $10^3$ TeV. 
In Fig.~\ref{JETp0_1TeV}
the results of the QGSJET simulations and the fits based on
Eq.(\ref{fitQG2}) are shown for the proton 
energies of 0.1 TeV and $10^3$ TeV.

The results of numerical simulations of the energy distribution
of secondary  pions  obtained with the SIBYLL code are well described
by the  function  
\be
\label{SIBYLL1}
\Phi_{\rm SIBYLL}=-B_\pi\left(\frac{1-x^\alpha}{1+r\,x^\alpha\,(1-x^\alpha)}\right)^4\,,
\ee
with the best fit parameters
\be
B_\pi=a+0.25\,,\quad \alpha=\frac{0.98}{\sqrt{a}}\,,\,r=\frac{2.6}{\sqrt{a}}\,,
\ee
where
\be
a=3.67+0.83\,L+0.075\,L^2\,, \quad L=\ln(E_p/1\,{\rm TeV})\,.
\ee

From Eqs.(\ref{SIBYLL1}) and (\ref{fit1}), and introducing an additional
term $(1-m_\pi/(x\,E_p))^{1/2}$, one obtains
\begin{eqnarray}
F_\pi(x,\,E_p)=4\alpha B_\pi x^{\alpha-1}\left(\frac{1-x^\alpha}
{1+rx^\alpha(1-x^\alpha)}\right)^4 \times\nonumber\\
\left(\frac{1}{1-x^\alpha}
+\frac{r\,(1-2x^\alpha)}{1+rx^\alpha(1-x^\alpha)}
\right)\left(1-\frac{m_\pi}{x\,E_p}\right)^{1/2}\,.
\label{SIBYLL2}
\end{eqnarray}

In the context of  different astrophysical applications,
the ultimate aim of this study is the analytical
description of gamma-rays and lepton from 
decays of unstable secondary particles
produced at proton-proton interactions. In this regard, in the case
of gamma-rays one has to consider, in addition to $\pi^0$-mesons,
also production and decay of $\eta$-mesons.
The analysis of the simulations show that the energy spectrum of
$\eta$-mesons is well described by the function
\be
\label{SIBYLL3}
F_\eta(x,\,E_p)=(0.55+0.028\,\ln x)\left(1-\frac{m_\eta}{x\,E_p}\right)F_\pi(x,\,E_p)
\ee
with the  condition  $F_\eta(x,\,E_p)=0$ at $x<m_\eta/E_p$.

\begin{figure*}[t]
\centering{%
\mbox{\includegraphics[width=0.35\textwidth,angle=-90,clip=]{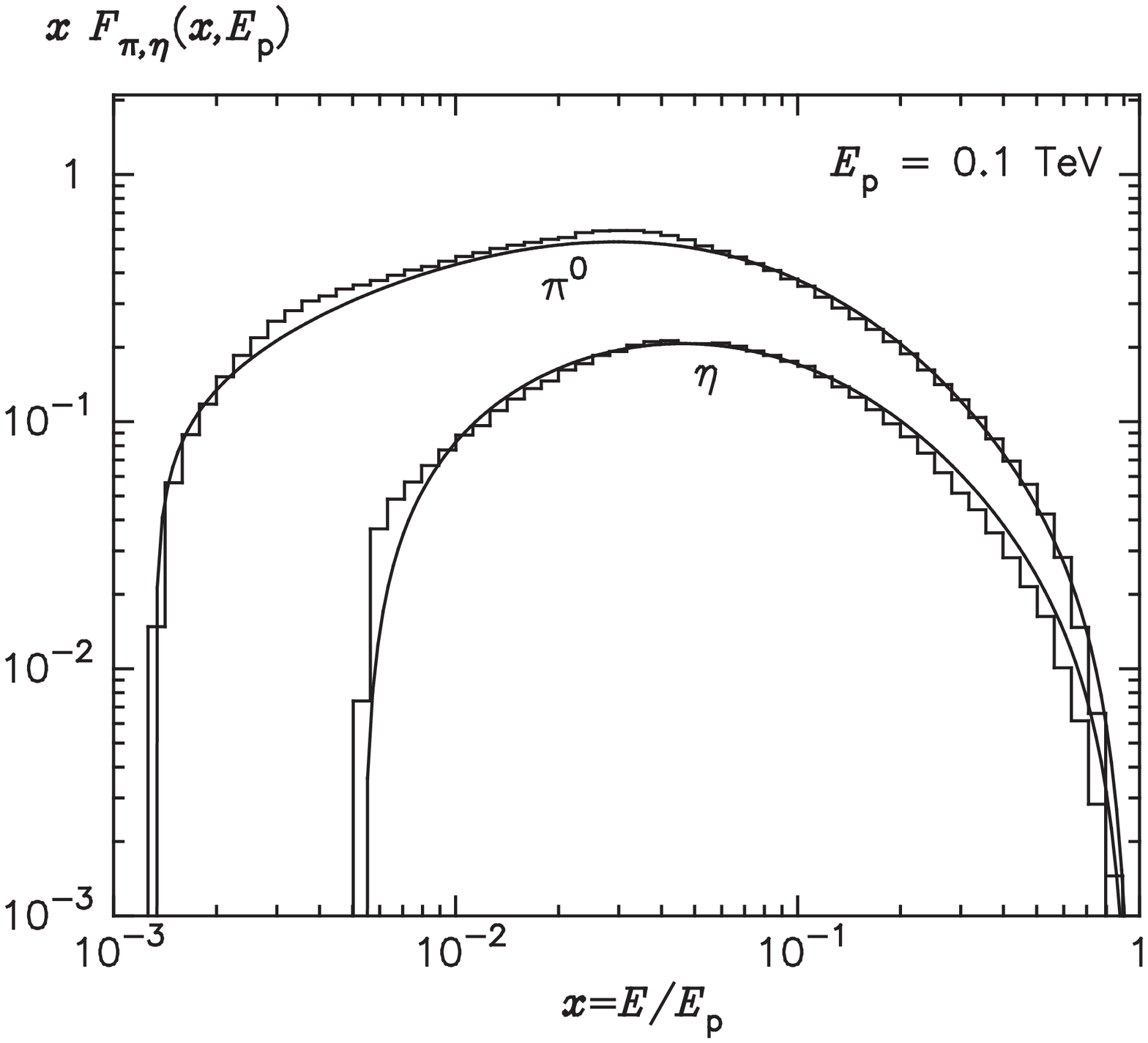}\quad
\includegraphics[width=0.35\textwidth,angle=-90,clip=]{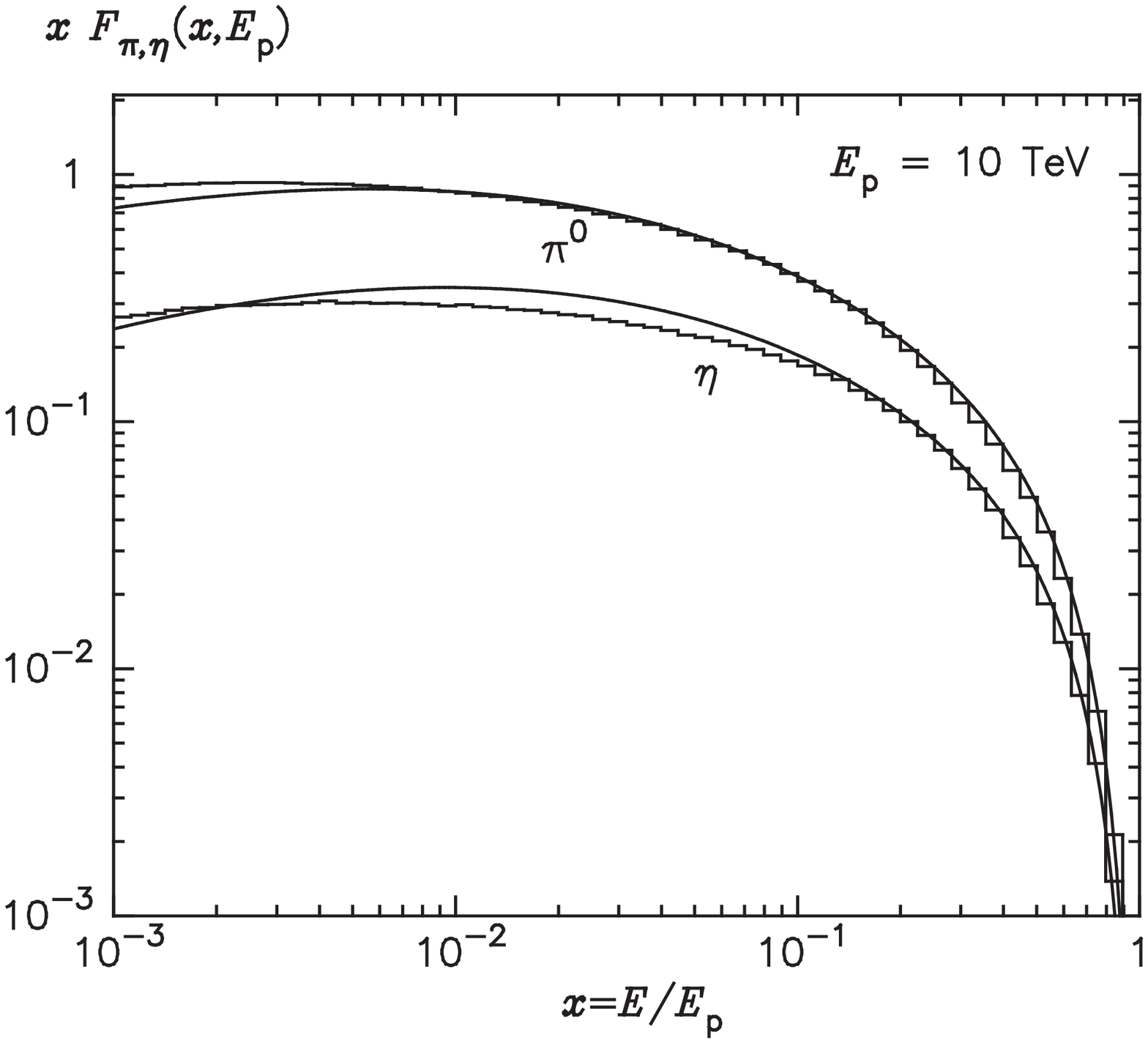}}
\mbox{\includegraphics[width=0.35\textwidth,angle=-90,clip=]{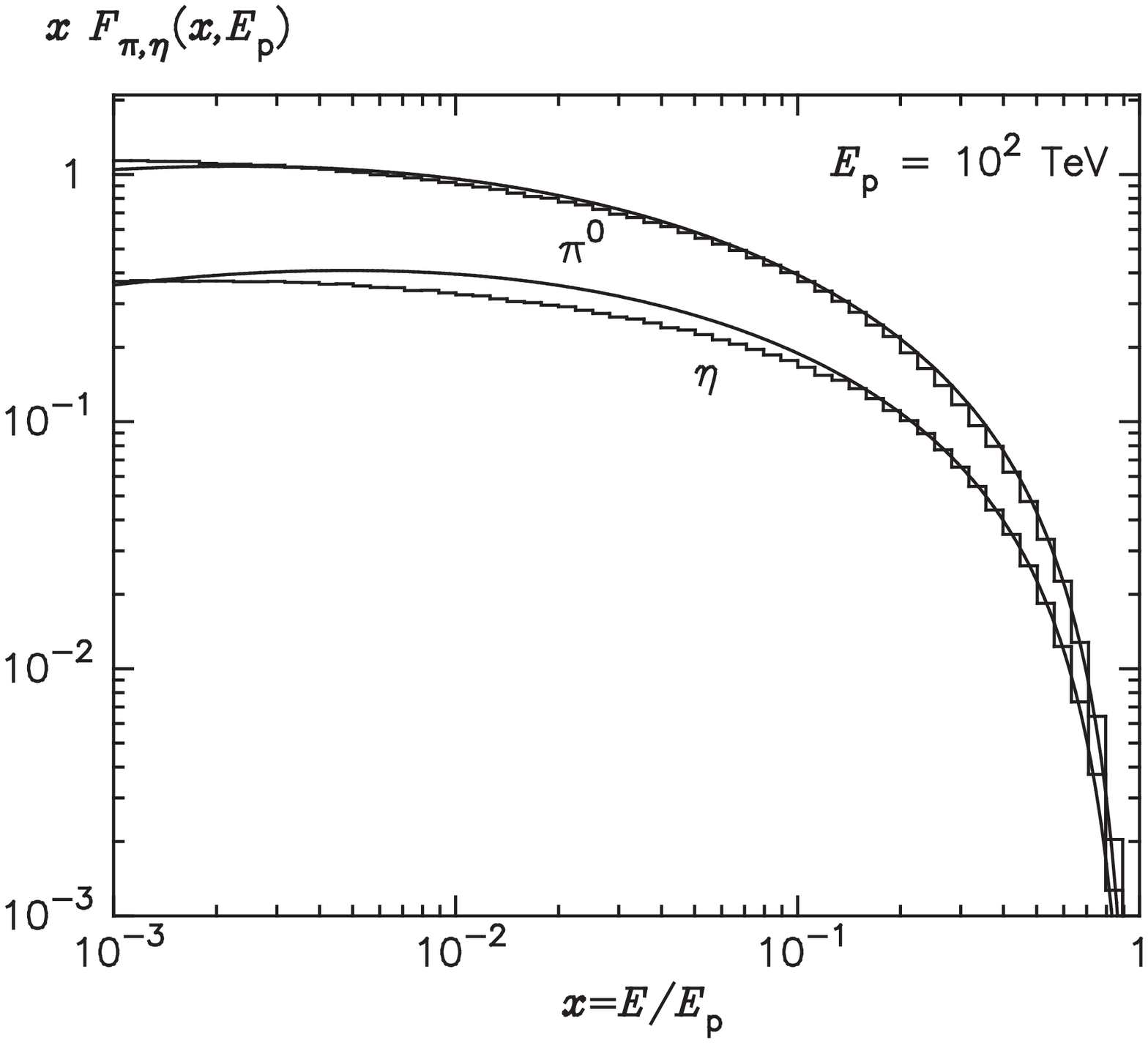}\quad
\includegraphics[width=0.35\textwidth,angle=-90,clip=]{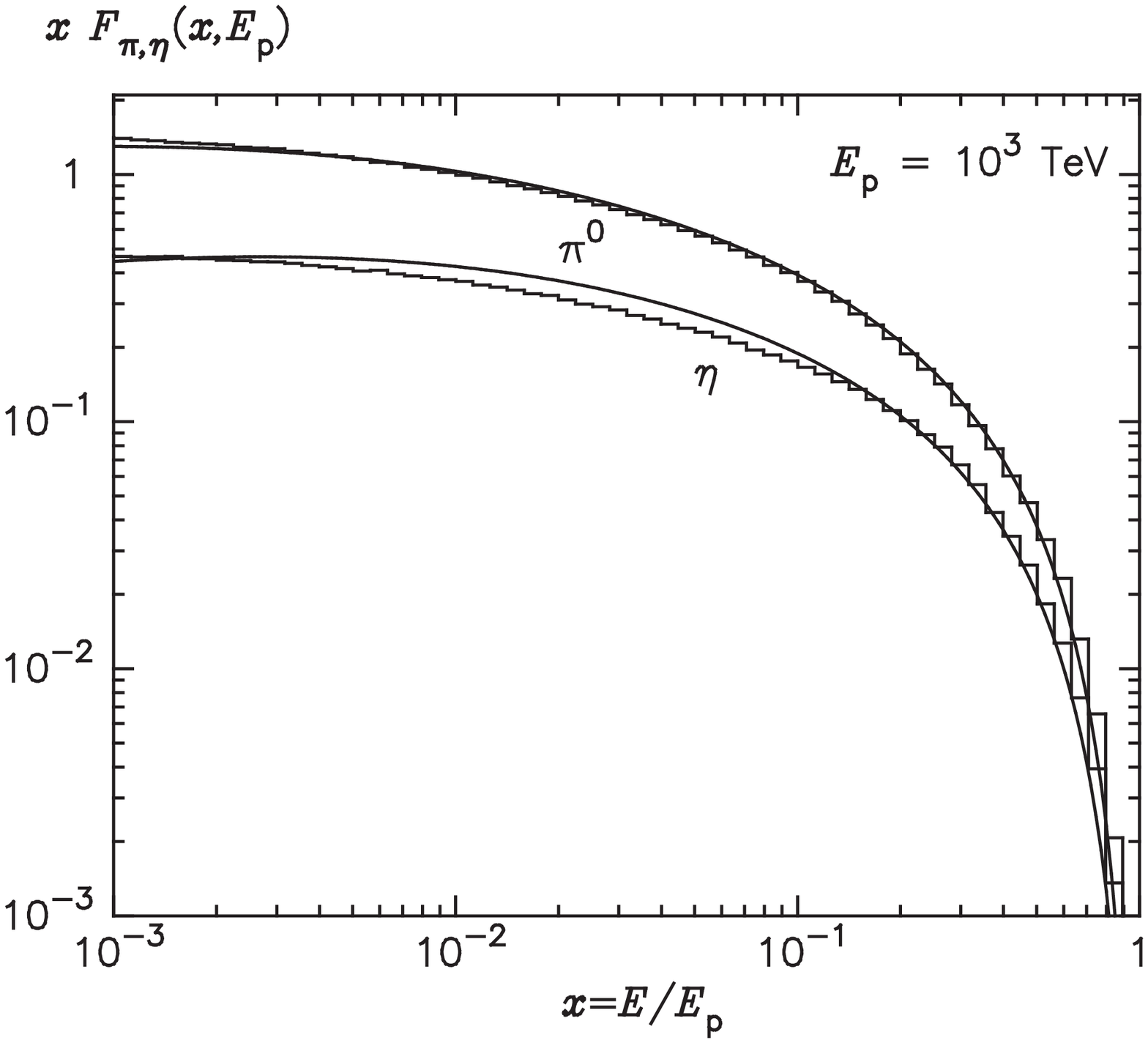}}
\caption{\small Energy spectra of $\pi$- and $\eta$-mesons from the
numerical simulations of the SIBYLL code (histograms) and from
the analytical presentations given by Eqs. (\ref{SIBYLL2}) and
(\ref{SIBYLL3}) for four energies of
primary protons: (a) 0.1~TeV, (b) 10~TeV, (c) 100~TeV, (d) 1000~TeV.}
\label{hist0_1TeV}
}
\end{figure*}

In Fig.~\ref{hist0_1TeV} we present the results of simulations
of energy distributions of $\pi$- and $\eta$-mesons
obtained with the SIBYLL code, together with the
analytical approximations given by Eqs. (\ref{SIBYLL2})
and (\ref{SIBYLL3}) for four energies of incident protons:
0.1, 10, 100, and 1000 TeV. Note that Eq.(\ref{SIBYLL2})
describes the spectra of pions at $x\agt10^{-3}$
with an accuracy better than 10 percent over the
energy range of protons from 0.1~TeV to $10^5\;{\rm TeV}$.
The approximation for the spectrum of $\eta$-mesons
is somewhat less accurate. However, since the
contribution of $\pi^0$-mesons to gamma-rays
dominates over the contribution from decays of
$\eta$-mesons, accuracy
of Eq. (\ref{SIBYLL3}) is quite acceptable.

\begin{figure}[h]
\centering{
\includegraphics[width=0.35\textwidth,angle=-90,clip=]{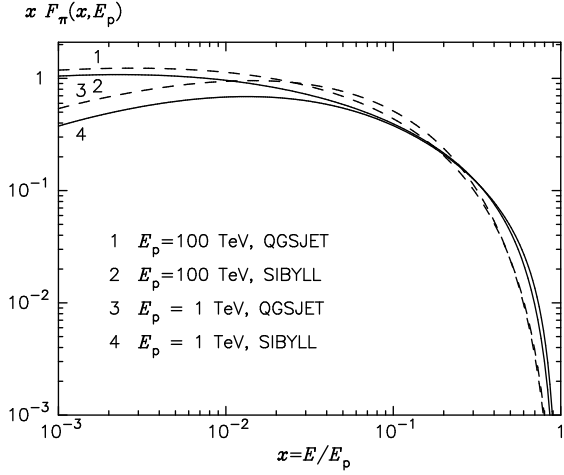}
\caption{\small Comparison of the energy distributions of secondary 
$\pi$-mesons  given  by Eq.(\ref{fitQG1}) (dashed curves) 
and Eq.(\ref{SIBYLL2}) (solid curves) 
calculated for two energies of parent protons $E_p=1$~TeV
and 100~TeV.}
\label{qgs_sib}
}
\end{figure}

The QGSJET and SIBYLL codes give quite similar, but not 
identical results. The differences  are  reflected in the 
corresponding analytical presentations. 
In Fig.\ref{qgs_sib} the energy spectra of pions  
described by Eqs.(\ref{fitQG1}) and (\ref{SIBYLL2}) are shown 
for two energies of protons. While these two approximations give similar
results  around $x \sim 0.1$, deviations at very small, $x \leq 10^{-2}$,
and very large, $x \geq 0.5$,  energies of secondary pions, are quite significant. 
Generally, the gamma-ray spectra from astrophysical objects are formed  at a
single \textit{p-p} collision (the so-called "thin target" scenario),
therefore the most important contribution comes from the region
$x=E_\pi/E_0 \sim  0.1$,  where these 
two approximations agree quite well with each other. 
On the other hand, for the proton distributions with 
distinct spectral features, in particular with sharp high-energy cutoffs, 
the region of $x \geq 0.1$ plays  important role in the formation of the
corresponding high-energy spectral tails of secondary products. 
Since it appears that in this range the
SIBYLL code describes the accelerator data somewhat better  
(Sergey Ostapchenko, private communication), in the following
sections we will use the distributions of secondary hadrons 
obtained with the SIBYLL code to  parameterize the 
energy spectra of  the final products of  \textit{p-p} interactions - 
gamma-rays, electrons and neutrinos.

\section{Decay of pions}
\label{pi-decay}
In this section we discuss the energy spectra of decay  products 
of neutral and charged ul\-tra\-re\-la\-ti\-vis\-tic $\pi$-mesons.

\subsection{\large $\pi^0 \to \gamma \gamma$}

The energy distribution of gamma-rays
from the decay of $\pi^0_{}$-mesons
with spectrum $J_{\pi}(E_{\pi})$ reads
\be
\label{SPG1}
Q_\gamma(E_\gamma)=2\int\limits_{E_\gamma}^\infty\!J_{\pi}(E_{\pi})\,
\frac{dE_{\pi}}{E_{\pi}}\,,
\ee
with total number of  photons
\be
\label{SPG2}
{\cal N}_\gamma=\int\limits_0^\infty\!Q_\gamma(E_\gamma)\,dE_\gamma=
2\int\limits_0^\infty\!dE_\gamma\int\limits_{E_\gamma}^\infty\frac{dE_{\pi}}{E_{\pi}}\,
J_{\pi}(E_{\pi})\,.
\ee
By changing the order of integration, and
integrating over $dE_\gamma$, one obtains an obvious relation
\be
\label{SPG3}
{\cal N}_\gamma=2\int\limits_{0}^\infty\!J_{\pi}(E_{\pi})\,dE_\pi=
2\,{\cal N}_\pi\, .
\ee
The total energy of gamma-rays also coincides with the total energy
of pions:
\be
\label{SPG4}
{\cal E}_\gamma=\int\limits_0^\infty\!E_\gamma\,Q_\gamma(E_\gamma)\,dE_\gamma
=\int\limits_0^\infty\!E_\pi\,J_\pi(E_\pi)\,dE_\pi={\cal E}_\pi\,.
\ee

Let's consider a simple example assuming that the spectrum
of pions has a power-law form with a low energy cutoff at $E_0$:
\be
\label{simple spectr}
J_{\pi}(E_{\pi})=\left\{
\begin{array}{cl}
A/(E_\pi)^{\alpha}\,,& \quad E_\pi>E_0\,,\\
0\,,&\quad E_\pi<E_0\,,
\end{array}
\right.
\ee
where $A$, $E_0$ and $\alpha$ are arbitrary constants.
For $\alpha>2$ the number of pions and their total
energy are equal
\be
{\cal N}_\pi=\frac{A}{(\alpha-1)E_0^{\alpha-1}}\,,\quad
{\cal E}_\pi=\frac{A}{(\alpha-2)E_0^{\alpha-2}} \, ,
\ee
while the energy spectrum of gamma-rays has the following form
\be
Q_\gamma(E_\gamma)=\left\{
\begin{array}{cl}
\ds\frac{2\,A}{\alpha E_\gamma^\alpha}\,,& \quad E_\gamma>E_0\,,\\[12pt]
\ds\frac{2\,A}{\alpha E_0^\alpha}\,,& \quad E_\gamma<E_0\,.
\end{array}
\right.
\ee
Above the energy $E_0$, the gamma-ray spectrum repeats the shape
of the pion spectrum; below $E_0$ the gamma-ray spectrum is energy-independent.

Note that in the region $E>E_0$ the number
of gamma-rays of given energy $E$ is less
than the number of pions of same energy. Indeed,
\be
R_\gamma\equiv\frac{Q_\gamma(E)}{J_{\pi}(E)}=\frac2\alpha<1\,,\quad \mbox{at}\;\; E>E_0\, .
\ee
On the other hand,  since the total number
of gamma-rays by a factor of 2 exceeds the total number of pions,
one should expect more gamma-rays at low energies.
Indeed, the numbers of gamma-rays above and below $E_0$ are
\begin{eqnarray*}
 {\cal N}_\gamma(E_\gamma>E_0)&=
&\frac{2\,A}{\alpha(\alpha-1)E_0^{\alpha-1}}\,,\\
{\cal N}_\gamma(E_\gamma<E_0)&=
&\frac{2\,A}{\alpha E_0^{\alpha-1}}\,.
\end{eqnarray*}
Although
${\cal N}_\gamma(E_\gamma>E_0)/{\cal N}_\pi=R_\gamma<1$,
the sum
\be
{\cal N}_\gamma(E_\gamma>E_0)+{\cal N}_\gamma(E_\gamma<E_0)=2\,{\cal N}_\pi\,.
\ee
One may say that the "deficit" of the number of high energy gamma-rays
is compensated by the "excess" of low energy gamma-rays.

\subsection{\large $\pi \to \mu \nu_\mu$}

In the rest frame of
the pion, the energy of secondary muons and neutrinos from the 
$\pi \to \mu \nu_\mu$ decays  
are\footnote{Hereafter we assume that the speed of light $c=1$.}
\be
\label{pi1}
E_\mu^0=\frac{m_\pi^2+m_\mu^2}{2\,m_\pi}\,,\quad
E_\nu^0=\frac{m_\pi^2-m_\mu^2}{2\,m_\pi}\,.
\ee
The momenta of secondaries in this system are
\be
p_\mu^0=p_\nu^0=E_\nu^0\,.
\ee
In the laboratory frame of coordinates (L-frame)
and for $E_\pi\gg m_\pi$
\be
E_{\nu,\max}=\frac1{m_\pi}\,(E_\pi E_\nu^0+p_\pi p_\nu^0)\approx
\lambda\,E_\pi\,,
\ee
where
\be
\lambda=1-m_\mu^2/m_\pi^2=0.427 \ .
\label{lambda}
\ee

Thus, the muonic neutrinos produced at the decay of an
ultrarelativistic pion of energy $E_\pi$ are distributed as
$dE_\nu /\lambda\,E_\pi$  ($0<E_\nu<\lambda\,E_\pi$).
Correspondingly, if the energy distribution of pions is
described by function $J_{\pi}(E_{\pi})$,
for neutrinos produced at the decay $\pi\to\mu\nu_\mu$
\be
\label{SPNU1}
Q_\nu(E_\nu)=\int\limits_{E_\nu/\lambda}^\infty\!J_{\pi}(E_{\pi})\,
\frac{dE_{\pi}}{\lambda\,E_{\pi}}\,.
\ee
For example, in the case of pion distribution
given by Eq.(\ref{simple spectr}), one has
\be
Q_\nu(E_\nu)=\left\{
\begin{array}{cl}
\ds\frac{\lambda^{\alpha-1}A}{\alpha E_\nu^\alpha}\,,& \quad E_\nu>\lambda E_0\,,\\[12pt]
\ds\frac{A}{\alpha\lambda E_0^\alpha}\,,& \quad E_\nu<\lambda E_0\,.
\end{array}
\right.
\ee
The ratio of neutrinos to pions of same energy is
\be
R_\nu=\frac{Q_\nu(E)}{J_{\pi}(E)}=\frac{\lambda^{\alpha-1}}\alpha<1
\,,\quad \mbox{at}\;\; E>E_0\,.
\ee
Note that the factor $\lambda^{\alpha-1}$ significantly reduces
the number of neutrinos relative to their counterpart
gamma-rays from decays of neutral pions.

\subsection{\large $\mu \to e \nu_e \nu_\mu$}

The treatment of this three-particle-decay channel is more complex.
Since $m_\mu\gg m_e$, in calculations we will neglect the mass of
electrons. Then the maximum energy of the muon
in the L-frame is
\be
E_{\mu,\max}=\frac1{m_\pi}\,(E_\pi E_\mu^0+p_\pi p_\mu^0)\approx
E_\pi\,.
\ee
Thus, the spectra of all particles produced at the decay
of muons will continue up to the energy $E_\pi$ (note that the neutrinos from the
decay of charge pions continue to $0.427 E_\pi$).
The formalism of calculations of energy distributions of secondary
products from decays of muons is described in \cite{Gaisser,Lipari}.

In $\pi \to \mu + \nu_\mu$ decays muons are fully polarized. In this case
the energy and angular distribution
of electrons (and muonic neutrinos)
in the rest frame of muon is described by the function
(see e.g. \cite{Okun})
\be
\label{distr e}
g_e(E'_e, \tilde\theta')=
\frac1{\pi m_\mu}\,\tau^2\left[3-2\tau \mp(1-2\tau)\cos\tilde\theta'\right]\,,
\ee
where $\tau={2E'_e}/{m_\mu}$, $\tilde\theta'$ is the angle between the momentum of electron
(neutrino) and the spin of muon. The signs $\mp$ of the second term in Eq.(\ref{distr e})
correspond to the decays of $\mu^+$ and $\mu^-$, respectively.

While the distribution
of muonic neutrinos is identical to the distribution of
electrons, the distribution $g_{\nu_e}(E'_{\nu_e}, \theta')$ of
electronic neutrinos has the following form:
\be
\label{distr nu}
g_{\nu_e}(E'_{\nu_e},\tilde\theta')=
\frac{6}{\pi m_\mu}\,\tau^2(1-\tau)\left(1\mp\cos\tilde\theta'\right)\,,
\ee
where $\tau={2E'_{\nu_e}}/{m_\mu}$
In the rest frame of the muon the maximum energy of each particle is $m_\mu/2$.
The functions given by Eqs.(\ref{distr e}) and
(\ref{distr nu}) are normalized as
\be
\int\!g(E',\tilde\theta')\,dE'\,d\Omega'=1\,.
\ee

Let's denote the angle between  the muon momentum  
in the rest frame of the pion and the electron momentum 
in the rest frame of muon as  $\theta'$.
Since the spin of  $\mu^-$ is parallel to the momentum,  
$\theta'=\tilde\theta'$, while for $\mu^+$ we have 
$\theta'=\pi-\tilde\theta'$ (the spin is anti-parallel to the momentum). 
Therefore the energy and angular distributions of electrons 
(expressed through $\theta'$), from  $\mu^+$ and  $\mu^-$  
decays have the same form:
\be
\label{electron}
g_e(E'_e,\theta')=
\frac1{\pi m_\mu}\,\tau^2\left[3+2\tau \,(1-2\tau)\cos\theta'\right]\,,
\ee
For the electronic neutrino we have 
\be
\label{neutrino}
g_{\nu_e}(E'_{\nu_e},\theta')=
\frac{6}{\pi m_\mu}\,\tau^2(1-\tau)\left(1+\cos\theta'\right)\,.
\ee

It is easy to obtain, after Lorentz transformations, 
the angular and energy distributions
of the decay products, $g_i(E, \theta)$, in the L-frame.  The  integration
of $g_i(E, \theta)$ over the solid angle  gives the 
corresponding energy distributions of particles
$f_i(E)$. These functions are derived in the Appendix.
Note that the energy distribution of leptons  
from the  $\pi^+$ and  $\pi^-$-meson decays are \textit{identical}
because of the CP invariance of weak interactions.
In this regard we should note that the statement of Ref.\cite{Mosk} 
about the "$e^\pm$ asymmetry"  is not correct.

For ultrarelativistic $\pi$-mesons the result can be presented
in the form of functions of variable
$x=E/E_\pi$, where $E$ is the energy of the lepton.
Let's denote  the probability of appearance of the variable $x$  
within the interval $(x,x+dx)$ as $dw=f(x)\,dx$.
The analytical integration gives the following distributions for
electrons and muonic neutrinos:
\begin{eqnarray}\label{f1}
&f_e(x)=f_{\nu_\mu^{(2)}}(x)=g_{\nu_\mu}(x)\,\Theta(x-r)+&\nonumber\\[4pt]
&(h^{(1)}_{\nu_\mu}(x)+h^{(2)}_{\nu_\mu}(x))\,\Theta(r-x)\,,&
\end{eqnarray}
where $r=1 - \lambda = (m_\mu/m_\pi)^2=0.573$,
\be
g_{\nu_\mu}(x)=\frac{3-2r}{9(1-r)^2}\,\left(9x^2-6\ln x-4x^3-5\right),
\ee
\be
h^{(1)}_{\nu_\mu}(x)=\frac{3-2r}{9(1-r)^2}\,\left(9r^2-6\ln r-4r^3-5\right),
\ee
\begin{eqnarray}\label{f2}
&h^{(2)}_{\nu_\mu}(x)=\ds\frac{(1+2r)(r-x)}{9r^2}\times&\nonumber\\[4pt]
&\left[9(r+x)-4(r^2+rx+x^2)\right] \ ,&
\end{eqnarray}
$\Theta$ is the Heaviside function ($\Theta(x)=1$ if $x \geq 0$, and
$\Theta(x)=0$ for $x < 0$).

For electronic neutrinos
\begin{eqnarray}\label{f3}
&f_{\nu_e}(x)=g_{\nu_e}(x)\,\Theta(x-r)+&\nonumber\\[4pt]
&(h^{(1)}_{\nu_e}(x)+h^{(2)}_{\nu_e}(x))\,\Theta(r-x)\,,&
\end{eqnarray}
where
\begin{eqnarray}\label{f4}
&g_{\nu_e}(x)=\ds\frac{2(1-x)}{3(1-r)^2}\times&\nonumber\\[4pt]
&\left[6(1-x)^2+r(5+5x-4x^2))+6r\ln x\right]\,,&
\end{eqnarray}
\begin{eqnarray}\label{f5}
&h^{(1)}_{\nu_e}(x)=\ds\frac2{3(1-r)^2}\times&\nonumber\\[4pt]
&\left[(1-r)(6-7r+11r^2-4r^3)+6r\ln r\right],&
\end{eqnarray}
\begin{eqnarray}\label{f6}
&h^{(2)}_{\nu_e}(x)=\ds\frac{2(r-x)}{3r^2}\times&\nonumber\\[4pt]
&\left(7r^2-4r^3+7xr-4xr^2-2x^2-4x^2 r\right).&
\end{eqnarray}
The functions are normalized
\be
\int\limits_0^1\!f_{\nu_\mu^{(2)}}(x)\,dx=\int\limits_0^1\!f_{\nu_e}(x)\,dx=1\,.
\ee

At $x=0$ one has $f_{\nu_\mu^{(2)}}(0)=2.214$ and
 $f_{\nu_e}(0)=2.367$, while at
$x \to 1 $ the functions behave as
\be
f_{\nu_\mu^{(2)}}(x)=\frac{2(3-2r)(1-x)^3}{3(1-r)^2}\,,\quad
f_{\nu_e}(x)=\frac{4(1-x)^3}{1-r}\,.
\ee

It is possible to obtain simple analytical presentations also for
the so-called Z-factors:
\begin{eqnarray}
&Z_{\nu_\mu}^{(2)} (\alpha)\equiv\int\limits_0^1\!x^{\alpha-1}f_{\nu_\mu^{(2)}}(x)\,dx=&\nonumber\\[4pt]
&\ds\frac{4[3-2r-r^\alpha(3-2r+\alpha-\alpha r)]}{\alpha^2(1-r)^2(\alpha+2)(\alpha+3)}\,,&
\end{eqnarray}
\begin{eqnarray}
&Z_{\nu_e} (\alpha)\equiv\int\limits_0^1\!x^{\alpha-1}f_{\nu_e}(x)\,dx=&\nonumber\\[4pt]
&\ds\frac{24[\alpha(1-r)-r(1-r^\alpha)]}{\alpha^2(1-r)^2(\alpha+1)(\alpha+2)(\alpha+3)}\,.&
\end{eqnarray}

For comparison, for $\gamma$-rays and neutrinos from the
$\pi \to \mu \nu_\mu$ decay, one has
\be
Z_\gamma=2 / \alpha \ ,
\ee
and
\be
Z_{\nu_\mu}^{(1)}=\lambda^{\alpha-1}/\alpha \, ,
\ee
where $\lambda$ is defined by Eq.(\ref{lambda}).

\begin{figure}[h]
\centering{%
\includegraphics[width=0.4\textwidth,angle=-90,clip=]{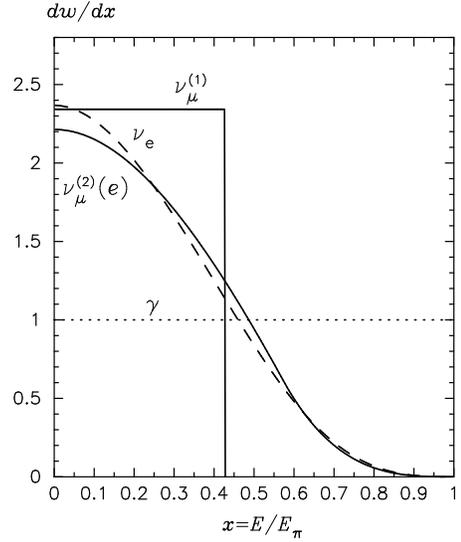}
\caption{\small Energy distributions of the secondary
products (photons, electrons, muonic and electronic neutrinos)
of decays of monoenergetic ultrarelativistic neutral
and charged pions. All distributions are normalized,
$\int_0^1\!dw=1$.}
\label{pi_mu_nu}
}
\end{figure}

In Fig.~\ref{pi_mu_nu} we show the functions
$dw/dx$ for  different decay products in the L-frame.
Note that the distributions of electrons and muonic neutrinos produced
at decays of muons simply coincides. In the same figure we show also
the distributions of photons (from the decay $\pi^0_{}\to2\gamma$
of neutral pions) and muonic neutrinos (from the decays $\pi\to\mu\nu_\mu$
of charged pions).
The curves corresponding to the distribution of muonic neutrinos are
marked by symbols $\nu_\mu^{(1)}$ for the neutrinos
from the direct decay $\pi\to\mu\nu_\mu$ and $\nu_\mu^{(2)}$
for the neutrinos from the decay $\mu\to e\nu_\mu\nu_e$.
All functions are normalized, $\int_0^1\!dw=1$.
%

\subsection{Energy spectra of decay products for arbitrary energy distributions of pions}

The spectra of secondary products of decays of pions of arbitrary
energy distributions $J_\pi(E_\pi)$ (the number of pions
in the energy interval $(E_\pi,\,E_\pi+dE_\pi)$ is equal
$dN_\pi(E_\pi)=J_\pi(E_\pi)\,dE_\pi$) are determined
by integration over the pion energy. In particular, for gamma-rays
\be
Q_\gamma(E_\gamma)
=2\int\limits_0^1\!J_\pi\!\left(\frac{E_\gamma}{x}\right)
\frac{dx}{x}\,.
\label{gamma-pi}
\ee
The factor 2 implies that at the decay of $\pi^0$-meson
two photons are produced. Similarly for the electrons and
electronic neutrinos
\be
\label{e-pi}
Q_e(E_e)=2\int\limits_0^1\!f_e(x)\,J_\pi\!\left(\frac{E_e}{x}\right)
\frac{dx}{x}\,,
\ee

\be
\label{nue-pi}
Q_{\nu_e}(E_{\nu_e})=2\int\limits_0^1\!f_{\nu_e}(x)\,J_\pi\!\left(\frac{E_{\nu_e}}{x}\right)
\frac{dx}{x}\,.
\ee

Here the factor 2 takes into account the contributions from
both $\pi^+$ and $\pi^-$  pions.

Finally, for muonic neutrinos we have
\[
Q_{\nu_\mu}(E_{\nu_\mu})=2\int\limits_0^1\!(f_{\nu_\mu^{(1)}}(x)+f_{\nu_\mu^{(2)}}(x))\,J_\pi\!
\left(\frac{E_{\nu_\mu}}{x}\right)\frac{dx}{x}=
\]
\be
\label{numu-pi}
\frac2\lambda\int\limits_0^\lambda\!J_\pi\!\left(\frac{E_{\nu_e}}{x}\right)\frac{dx}{x}+
2\int\limits_0^1\!f_{\nu_\mu^{(2)}}(x)\,J_\pi\!
\left(\frac{E_{\nu_\mu}}{x}\right)\frac{dx}{x}\,.
\ee
The first and second terms of Eq.(\ref{numu-pi})
describe the muonic neutrinos produced through the
direct decay $\pi\to\mu\nu_\mu$ of the pion and
the decay of the secondary muon, respectively.
Note that  in Eqs.(\ref{e-pi}) -- (\ref{numu-pi}) we do not distinguish between electrons and positrons , as well as
between neutrinos and antineutrinos. In fact, in $p-p$ interactions
the number of $\pi^+$- mesons slightly exceeds the number of
$\pi^-$ even at energies far from the threshold. However, this effect
is less than the accuracy
of both the measurements and our analytical approximations, therefore we will
adopt $\nu =\bar \nu$ and $e^+ = e^-$.

%
\begin{figure}[h]
\centering{%
\includegraphics[width=0.4\textwidth,angle=-90,clip=]{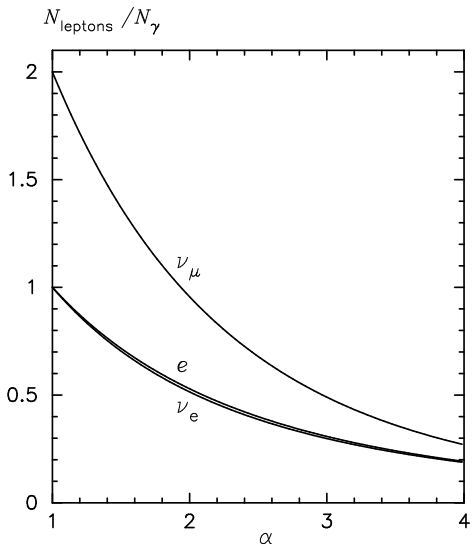}
\caption{\small The ratio of the number of leptons to photons of
same energy for power-law distributions of pions with spectral index
$\alpha$. It is assumed that
$\pi^0$, $\pi^+$ and $\pi^-$ have identical distributions.}
\label{nutogam}
}
\end{figure}

It is of practical interest to compare the spectra of the secondary particles for
power-law distributions of pions. Namely, let assume that
the energy distributions of $\pi^0_{}$, $\pi^-_{}$ and
$\pi^+_{}$ mesons are distributed by Eq.(\ref{simple spectr})
with an arbitrary spectral index $\alpha$. Then in the energy region
$E>E_0$ the energy spectra of photons and leptons
are power-law with the same spectral index $\alpha$. The ratio
of leptons to photons of same energy depends only on $\alpha$.
For the muonic neutrinos, electrons, and electronic neutrinos
these ratios are
\be
\label{nu-mu to gam}
\frac{N_{\nu_\mu}}{N_\gamma}=\lambda^{\alpha-1}+
\alpha\int\limits_0^1\!f_e(x)\,x^{\alpha-1}\,dx\,,
\ee
\be
\label{e to gam}
\frac{N_e}{N_\gamma}=\alpha\int\limits_0^1\!f_e(x)\,x^{\alpha-1}\,dx\,,
\ee
\be
\label{nu-e to gam}
\frac{N_{\nu_e}}{N_\gamma}=\alpha\int\limits_0^1\!f_{\nu_e}(x)\,x^{\alpha-1}\,dx\,.
\ee

The ratios given by Eqs.(\ref{nu-mu to gam}), (\ref{e to gam})
and (\ref{nu-e to gam}) as functions of $\alpha$ are shown in
Fig.~\ref{nutogam}. Note that only in the case of
$\alpha=1$ the ratio $N_{\rm leptons}/N_\gamma$
coincides with the ratio of the total number of leptons to
gamma-rays. With increase of $\alpha$
the lepton/photon ratio decreases. For example, for
$\alpha=2$ one has $N_{\nu_\mu}/N_\gamma=0.96$, $N_e/N_\gamma=0.53$,
$N_{\nu_e}/N_\gamma=0.52$.
For $\alpha=2.5$
these ratios are 0.68, 0.40, 0.39, respectively. Finally, note that the spectra of
electrons and electronic neutrinos are very similar, thus the corresponding curves
in Fig.\ref{nutogam} practically coincide.

\section{Energy spectra of photons and leptons produced at p-p collisions}
\label{spectra}
For calculations of  the energy spectra of gamma-rays, neutrinos
and electrons produced at \textit{p-p}
interactions one should substitute the
distribution of pions $F_\pi(x,\,E_p)$ given
by Eq. (\ref{SIBYLL2})
into Eqs.(\ref{gamma-pi}) -- (\ref{numu-pi}):
\be
\label{change}
J_{\pi}(E_{\pi})\,dE_{\pi} \to F_\pi(x,\,E_p)\,dx=
F_\pi\!\left(\frac{E_{\pi}}{E_p},\,E_p\right)\, \frac{dE_{\pi}}{E_p} \, .
\ee
The incident
proton energy $E_p$ enters this equation as a free parameter.

\subsection{Spectra of gamma-rays}

For calculations of gamma-ray spectra one  should
include the contribution also from $\eta$-mesons.
The main modes of decay of $\eta$-mesons
are \cite{ParticleData}:
$\eta\to 2\,\gamma$ (39.4\%); $\eta\to 3\,\pi^0_{}$ (32.5\%);
$\eta\to \pi^+_{}\pi^-_{}\pi^0_{}$ (22.6\%) and
$\eta\to \pi^+_{}\pi^-_{}\gamma$ (5\%). Approximately 3.2
gamma-ray photons  are produced per decay of $\eta$-meson; 
the energy transferred to all gamma-rays is $0.81\,E_\eta$.

The spectrum of gamma-rays produced in the direct
$\eta\to 2\,\gamma$ decay is calculated similarly to the
decay of $\pi^0_{}$-mesons.
For accurate calculation of gamma-ray spectra
from the decay chains $\eta\to 3\,\pi^0_{}\to6\gamma$
and $\eta\to\pi^+_{}\pi^-_{}\pi^0_{}\to2\gamma$,
one has to know, strictly speaking, the function of energy distribution
of pions $g(E_\pi^0)$ in the rest frame of the $\eta$-meson,
and then  calculate, through Lorentz transformations, the distribution
of pions in the L-frame. However, the calculations with model
functions show that the spectrum of gamma-rays
weakly depends
on the specific form of $g(E_\pi^0)$; for different extreme
forms of function $g(E_\pi^0)$ the results vary within 5 percent.

The maximum energy of pions in the rest frame of $\eta$-meson
is $E^0_{\pi\max}=(m_\eta^2-3\,m_\pi^2)/(2\,m_\eta)$, while in the L-frame
$
E_{\pi\max}=\left(E^0_{\pi\max}+p^0_{\pi\,\max}\right)E_\eta/m_\eta\approx
0.73\,E_\eta $.
Therefore the spectrum of gamma-rays from the decay chain of $\eta$-mesons
breaks at $E_\gamma=0.73\,E_p$.

\begin{figure*}[t]
\begin{center}
\mbox{\includegraphics[width=0.4\textwidth,angle=-90,clip=]{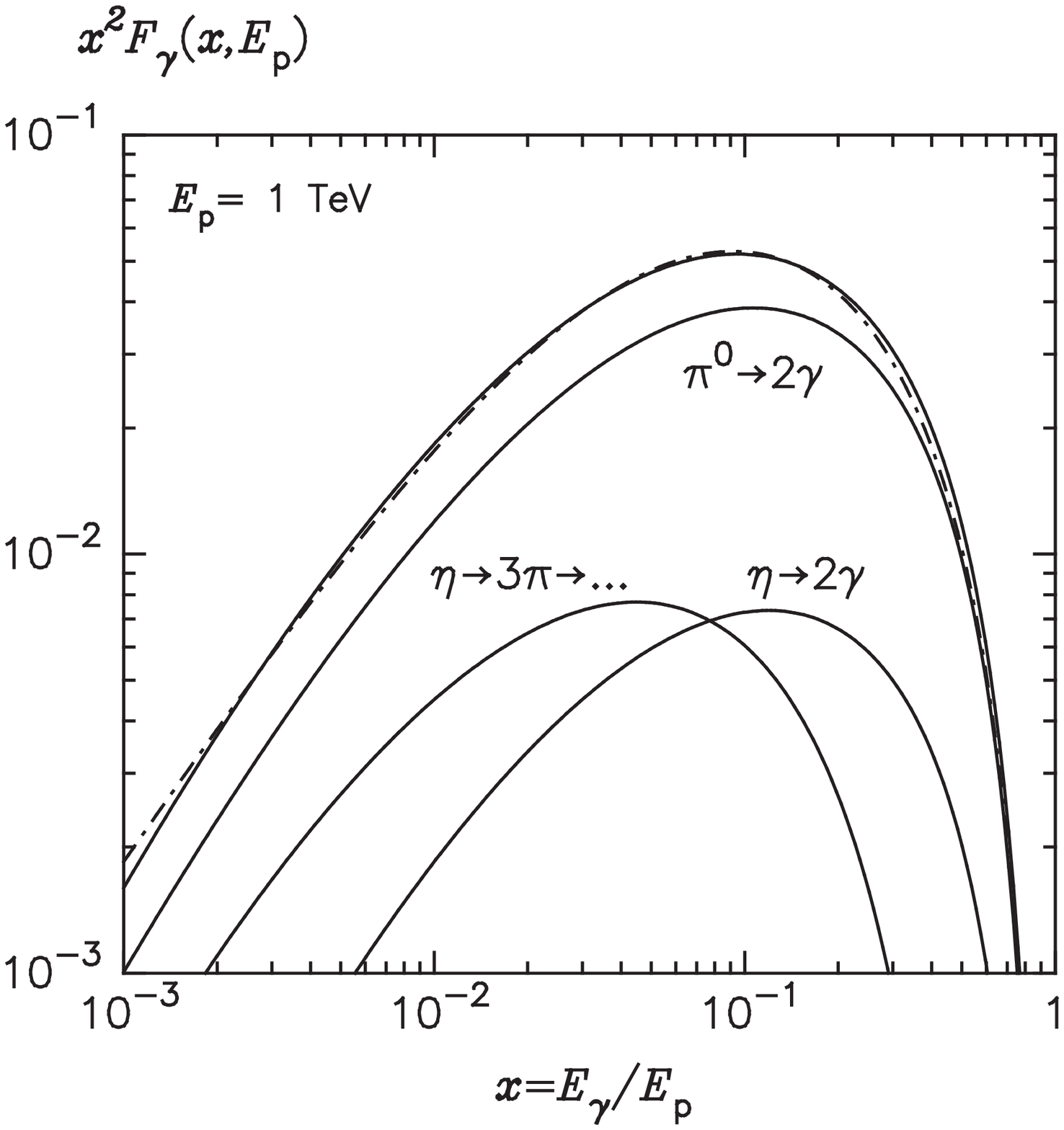}\quad
\includegraphics[width=0.4\textwidth,angle=-90,clip=]{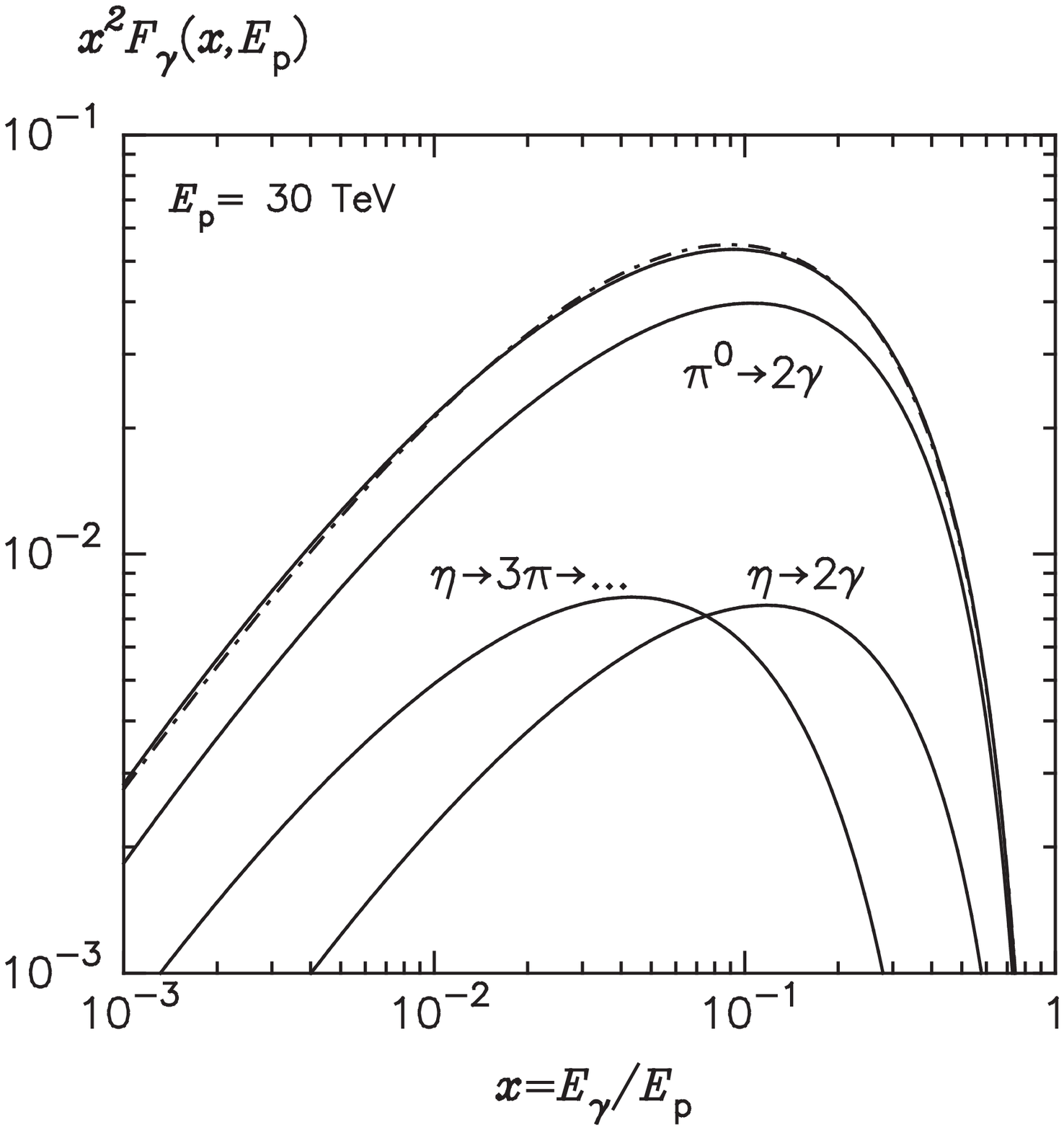}}
\mbox{\includegraphics[width=0.4\textwidth,angle=-90,clip=]{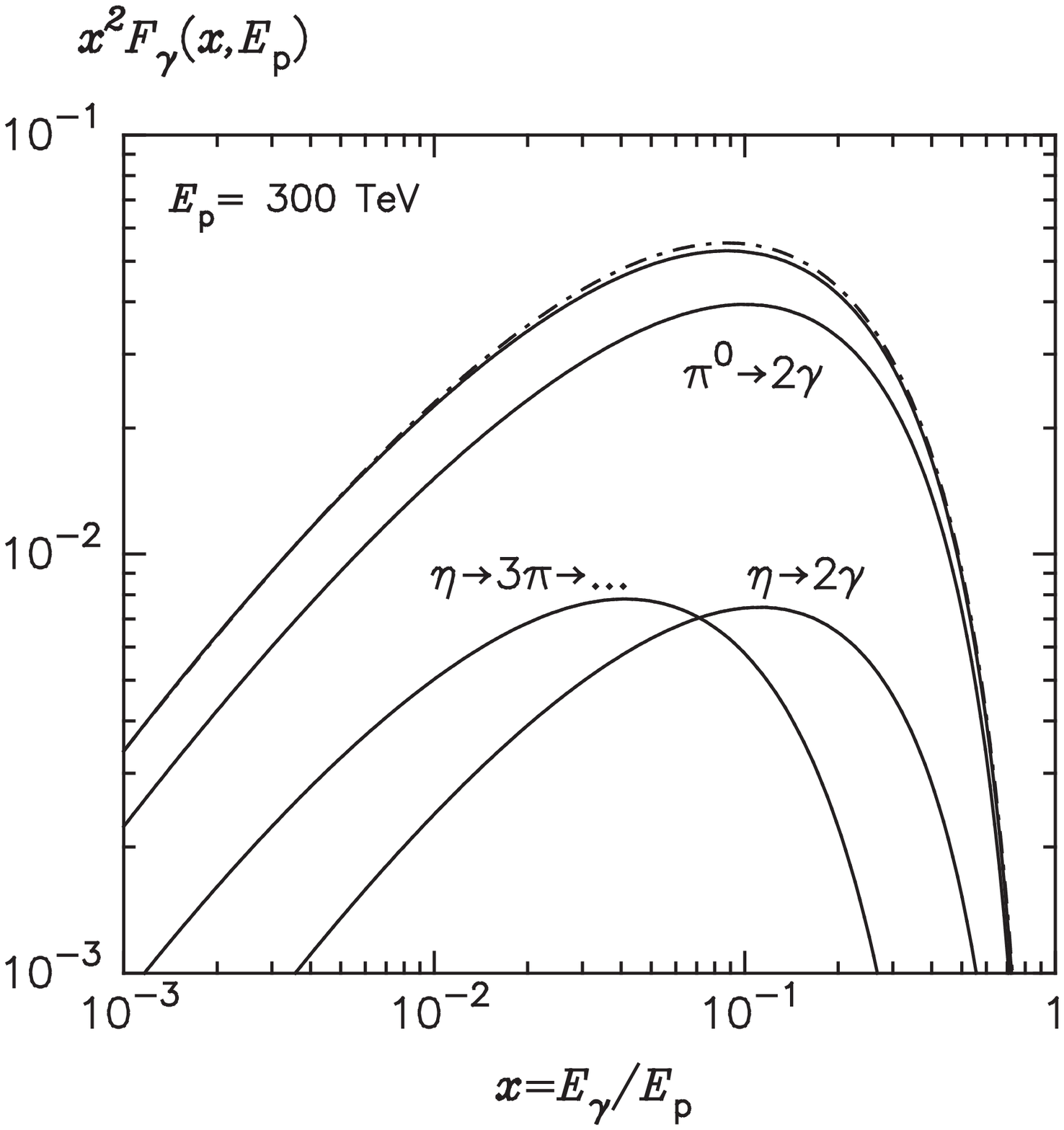}\quad
\includegraphics[width=0.4\textwidth,angle=-90,clip=]{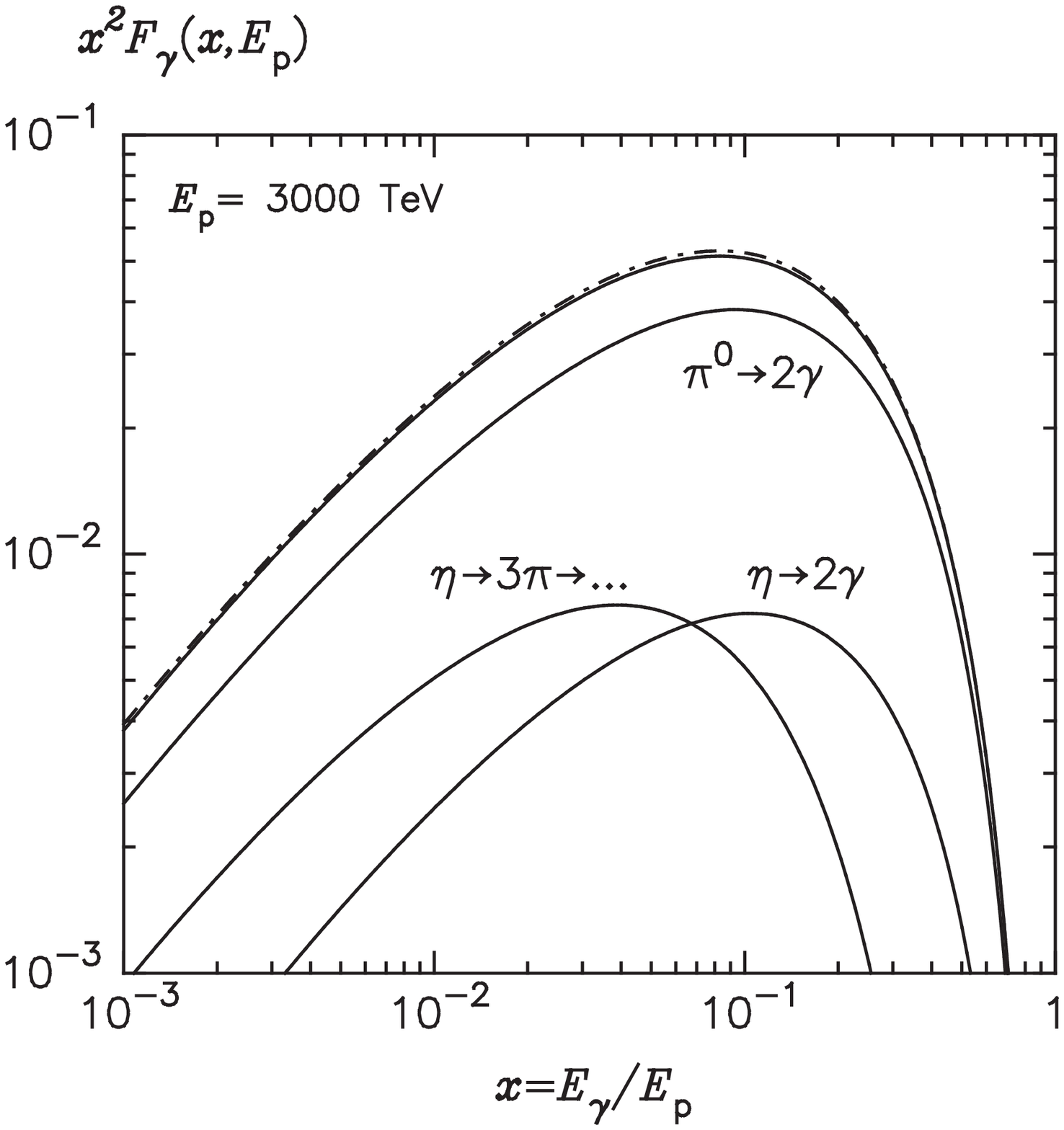}}
\caption{\small Energy spectra of gamma-rays from pp interactions
corresponding to four different energies of incident protons: (a)~1~TeV,
(b)~30~TeV, (c)~300~TeV, (d)~3000~TeV. Solid lines - numerical calculations based on
simulations of production of $\pi$ and $\eta$ mesons using the
SIBYLL code, dot-dashed line - analytical presentation given by
Eq.(\ref{fit_gamma}). The partial contributions from $\pi^0$ and
$\eta$ meson decays are also shown.} \label{p1gam}
\end{center}
\end{figure*}

The spectra of gamma-rays calculated for both $\pi^0$ and $\eta$-meson
decay channels are shown in Fig.~\ref{p1gam}. The total spectrum of gamma-rays,
based on the simulations of energy distributions
of $\pi$ and $\eta$-mesons by the SIBYLL code, can be presented in the
following simple analytical form
\begin{eqnarray}
F_\gamma(x,\,E_p)=
B_\gamma\,\frac d{dx}\left[\ln(x)\left(\frac{1-x^{\beta_\gamma}}
{1+k_\gamma x^{\beta_\gamma}(1-x^{\beta_\gamma})}\right)^4\right]=\nonumber\\[5pt]
B_\gamma\,\frac{\ln(x)}{x}\left(\frac{1-x^{\beta_\gamma}}
{1+k_\gamma x^{\beta_\gamma}(1-x^{\beta_\gamma})}\right)^4\times\qquad\nonumber\\[8pt]
\left[\frac1{\ln(x)}-\frac{4\beta_\gamma x^{\beta_\gamma}}{1-x^{\beta_\gamma}}
-\frac{4k_\gamma\beta_\gamma x^{\beta_\gamma}(1-2x^{\beta_\gamma})}
{1+k_\gamma x^{\beta_\gamma}(1-x^{\beta_\gamma})}
\right],\qquad&
\label{fit_gamma}
\end{eqnarray}
where $x=E_\gamma/E_p$. The function $F_\gamma(x,\,E_p)$ implies
the number of photons in the interval $(x,x+dx)$ per collision.
The parameters $B_\gamma$, $\beta_\gamma$, and $k_\gamma$
depend only on the energy of proton. The best least squares fits to the
numerical calculations of the spectra in the energy range of
primary protons $0.1\,{\rm TeV}\le E_p\le 10^5\,{\rm TeV}$ give
\begin{eqnarray}
B_\gamma&=&1.30+0.14\,L+0.011\,L^2\,,\\[4pt]
\beta_\gamma&=&\frac1{1.79+0.11\,L+0.008\,L^2}\,,\\[4pt]
k_\gamma&=&\frac1{0.801+0.049\,L+0.014\,L^2}\,,
\end{eqnarray}
where $L=\ln(E_p/1\,{\rm TeV})$. Note that around
$x\sim0.1$ the contribution from $\eta$-mesons is about 25 percent.
In the most important region of $x\agt10^{-3}$ Eq.(\ref{fit_gamma})
describes the results of numerical calculations with an
accuracy better than a few percent  provided that the energy
of gamma-rays $E_\gamma\agt1\,{\rm GeV}$. At low energies, 
this analytical presentation does not provide adequate
accuracy, therefore it cannot be used 
for the estimates of the total number of gamma-rays.

\begin{figure}[h]
\centering{%
\includegraphics[width=0.35\textwidth,angle=-90,clip=]{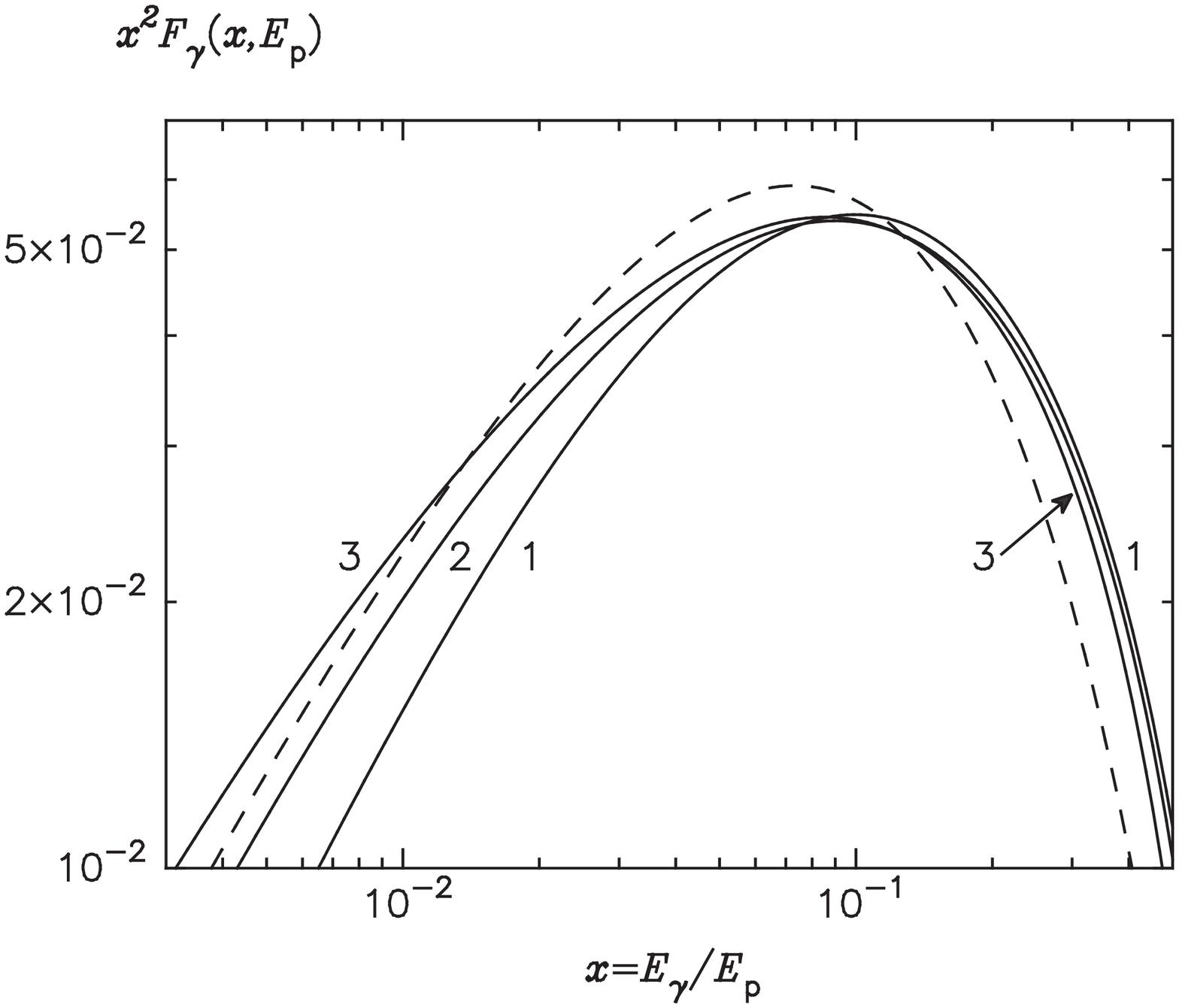}
\caption{\small Energy spectra of gamma-rays described by
Eq.(\ref{fit_gamma}) for three energies of incident protons:
0.1~TeV (curve 1), 100~TeV (curve 2) and 1000 TeV (curve 3). The dashed curve corresponds
to the Hillas parameterization of the spectra
obtained for proton energies of several tens of TeV.}
\label{g_for_diff_E}
}
\end{figure}

The numerical calculations of energy spectra of gamma-rays
from decays of $\pi^0$ and $\eta$ mesons
for four energies of incident protons, 1, 30, 300, and 3000 TeV, are shown
in Fig.~\ref{p1gam} together with the analytical presentations given by
Eq.(\ref{fit_gamma}). The numerical calculations as well as the analytical
presentations of the gamma-ray spectra shown in Fig.~\ref{p1gam}
are based on the simulations obtained with
the SIBYLL code. Recently Hillas \cite{Hillas}
suggested a simple  parameterization for gamma-ray spectrum
based on simulations of \textit{p-p} interactions at proton energies 
of a few tens of TeV using the QGSJET code:
$F_\gamma(x)=3.06\exp\left(-9.47\,x^{0.75}\right)$.
In Fig.~\ref{g_for_diff_E} we show the curve
based on this parameterization, together with
gamma-ray spectra using  Eq.(\ref{fit_gamma})
calculated for  proton energies 0.1, 100 and 1000~TeV.
While there is a general good  agreement between the curves shown
in Fig.~\ref{g_for_diff_E}, especially around $x \sim 0.1$, the parameterization
of Hillas gives somewhat  steeper spectra at $x \geq 0.1$
The difference basically comes from the different interaction codes used 
for parameterizations of gamma-ray spectra 
(compare the curves in  Fig.\ref{qgs_sib}).  While 
the Hillas parameterization (dashed line) is based on the 
QGSJET code,  the calculations for gamma-ray
spectra shown  in  Fig.~\ref{g_for_diff_E} by solid lines 
correspond to  the SIBYLL code.

%
Eqs.(\ref{f1}) -- (\ref{f6})

%

\subsection{Energy spectra of leptons}

The calculations of  energy distributions of electrons and neutrinos from 
\textit{p-p} have been performed numerically,   substituting Eq.(\ref{change}) into
Eqs. (\ref{e-pi}), (\ref{nue-pi}), and (\ref{numu-pi}). 
The spectra of electrons from the $\pi\to\mu\,\nu_\mu$ decays
is well described by the following  function
\be
\label{LEP1}
F_e(x,\,E_p)=B_e\,\frac{(1+k_e(\ln x)^2)^3}{x\,(1+0.3/x^{\beta_e})}\,(-\ln(x))^5\,,
\ee
where 
\begin{eqnarray}
B_e&=&\frac1{69.5+2.65\,L+0.3\,L^2}\,,\\[4pt]
\beta_e&=&\frac1{(0.201+0.062\,L+0.00042\,L^2)^{1/4}}\,,\\[4pt]
k_e&=&\frac{0.279+0.141\,L+0.0172\,L^2}{0.3+(2.3+L)^2}\,,
\end{eqnarray}
Here  $x=E_e/E_\pi$ and $L=\ln(E_p/1\,{\rm TeV})$.

The spectrum of muonic neutrino from the decay of muon,
$F_{\nu_\mu^{(2)}}(x,\,E_p)$,  is described  by the same function;
in this case $x=E_{\nu_\mu}/E_p$.

The spectrum of muonic neutrino produced through the direct
decays $\pi\to \mu\,\nu_\mu$ can be described as
\begin{eqnarray}
F_{\nu_\mu^{(1)}}(x,\,E_p)=
B'\,\frac{\ln(y)}{y}\left(\frac{1-y^{\beta'}}
{1+k'y^{\beta'}(1-y^{\beta'})}\right)^4\times\nonumber\\[6pt]
\left[\frac1{\ln(y)}-\frac{4\beta' y^{\beta'}}{1-y^{\beta'}}
-\frac{4k'\beta' y^{\beta'}(1-2y^{\beta'})}{1+k'y^{\beta'}(1-y^{\beta'})}
\right],
\label{LEP3}
\end{eqnarray}
where $x=E_{\nu_\mu}/E_p$, $y=x/0.427$,
\begin{eqnarray}
B'=1.75+0.204\,L+0.010\,L^2\,,\\[4pt]
\beta'=\frac1{1.67+0.111\,L+0.0038L^2}\,,\\[4pt]
k'=1.07-0.086\,L+0.002\,L^2\,.
\end{eqnarray}
The spectrum of $F_{\nu_\mu^{(1)}}$ sharply cutoffs at $x=0.427$.
The total spectrum of muonic neutrinos
$F_{\nu_\mu}=F_{\nu_\mu^{(1)}}+ F_{\nu_\mu^{(2)}}$, where
$F_{\nu_\mu^{(1)}}=F_{e}$. Also, with good accuracy
(less than 5\%) $F_{\nu_e} \approx F_{e}$.

\begin{figure}[h]
\centering{%
\includegraphics[width=0.35\textwidth,angle=-90,clip=]{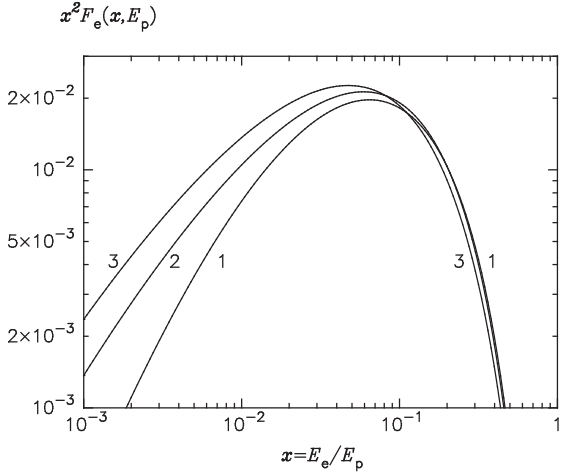}
\caption{\small Energy spectra of electrons described by
Eq.(\ref{LEP1}) for three energies of incident protons:
0.1~TeV (curve 1), 100~TeV (curve 2) and 1000~TeV (curve 3).
}
\label{e_for_diff_E}
}
\end{figure}

\begin{figure}[h]
\centering{%
\includegraphics[width=0.35\textwidth,angle=-90,clip=]{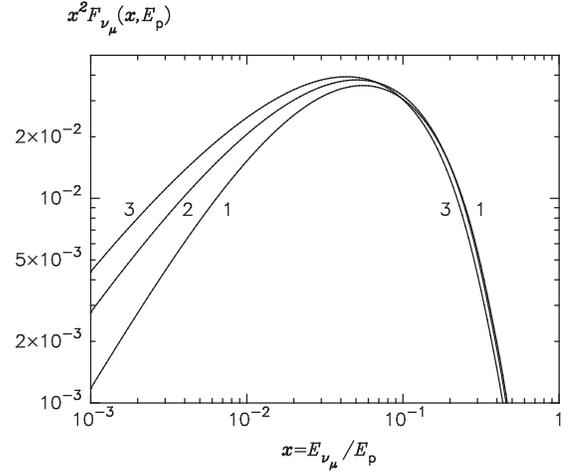}
\caption{\small \small Energy spectra of all muonic neutrinos described by
Eq.(\ref{LEP1}) and (\ref{LEP3}) for three energies of incident protons:
0.1~TeV (curve 1), 100~TeV (curve 2) and 1000~TeV (curve 3).
}
\label{numu_for_diff_E}
}
\end{figure}

The spectra of electrons and muonic neutrinos for different energies of
protons (0.1, 100, and 1000~TeV) are shown in Figs.~\ref{e_for_diff_E} and
\ref{numu_for_diff_E}, respectively.

Although  $\pi^\pm$-decays  strongly 
dominate in the  lepton production (see e.g. \cite{Gaisser}),
the contribution through  decays of other unstable secondaries, 
first of all K-mesons, is not negligible. 
In particular,  calculations based on simulations of all 
relevant  channels using the SIBYLL  code show that the 
total yield of electrons and neutrinos exceeds Eq.(\ref{LEP1}) by 
approximately 10 \%  at $x \sim 0.1$.  At very small values of $x$  the difference 
increases and  could be as large as 20 \% at $x \ll 0.1 $;  
at very large values of x, $x \geq  0.5$,   it is reduced to 5 \%. 
Thus,  for  power-law energy  distributions of protons,    
Eqs.(\ref{LEP1})  and (\ref{LEP3}) underestimate 
the flux of neutrinos and electrons  by a factor of 1.1. 
For production of gamma-rays, the simulations 
performed with the SIBYLL code show 
that the combined contribution  from  $\pi^0$ and $\eta$  
meson decays exceeds 95 \% of the total gamma-ray production rate. 

\begin{figure*}[t]
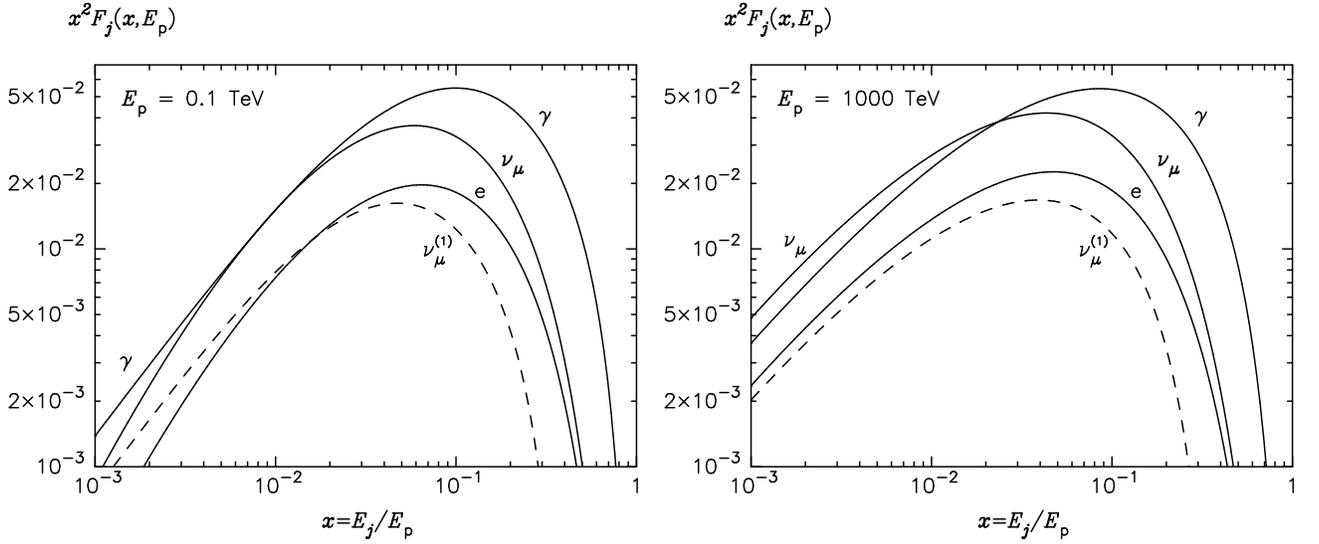

\centering{%
\mbox{%
\includegraphics[width=0.4\textwidth,angle=-90,clip=]{10a.eps}\quad
\includegraphics[width=0.4\textwidth,angle=-90,clip=]{10b.eps}
}
\caption{\small Energy spectra of all decay products produced at
\textit{p-p} interactions for two energies of incident protons: (a)~0.1~TeV and (b)~1000~TeV.
The spectrum of electronic neutrinos is not shown because it practically coincides
with the spectrum of electrons.}
\label{all_part01}
}
\end{figure*}

To compare the contributions of the final products of decays, in Fig.~\ref{all_part01}
we show the spectra of gamma-rays, electrons and muonic neutrinos for two energies of
incident protons, $E_p=0.1$ and 1000 TeV.
It is interesting that for significantly different energies of primary protons
the spectra of all secondary products are rather similar, although not
identical. This explains why for broad and smooth (e.g. power-law type)
energy distribution of protons the so-called delta-functional approach of
calculating gamma-ray spectra gives quite accurate results (e.g. \cite{AhAt00,Torres}).
However for proton spectra with distinct spectral features, like
sharp pileups or cutoffs, this method may lead  
to wrong conclusions concerning, in particular, the predictions of 
the energy spectra of gamma-rays and neutrinos.

%

%

%

\section{Production rates and spectra of photons and leptons for
wide  energy distributions of protons}

The simple analytical approximations presented in the previous section
for the energy spectra of secondary particles
produced by a proton of fixed  energy make the
calculations of production rates and spectra of gamma-rays and leptons
for an arbitrary energy distribution of protons quite simple.
Below we assume that the gas density as well as the
magnetic of the ambient medium are sufficiently low, so all secondary products
decay before interacting with the gas and the magnetic field.

Let denote by 
\be
\label{proton}
dN_p=J_p(E_p)\,dE_p
\ee
the number of protons in a unite volume in the energy interval
$(E_p,\,E_p+dE_p)$. Below for the function
$J_p(E_p)$ we will use  units ${\rm cm}^{-3}\,{\rm TeV}^{-1} $. 
Then the function  
\begin{eqnarray}
&\Phi_\gamma(E_\gamma)\equiv\ds\frac{dN_\gamma}{dE_\gamma}=&\nonumber\\[5pt]
&\ds c\,n_{H}^{}\int\limits_{E_\gamma}^\infty\! \sigma_{\rm inel}(E_p)\,
J_p(E_p)\,F_\gamma\!\left(\frac{E_\gamma}{E_p},\,E_p\right)\frac{dE_p}{E_p}\,.&
\label{Phi g1}
\end{eqnarray}
describes the gamma-ray production rate in the energy interval
$(E_\gamma,\,E_\gamma+dE_\gamma)$, 
where $n_{H}^{}$ is the density of the ambient hydrogen gas,
$\sigma_{\rm inel}(E_p)$ is the cross-section of inelastic \textit{p-p} interactions,
and $c$ is the speed of light. The function $F_\gamma(x,\,E_p)$ is defined by
Eq.(\ref{fit_gamma}). For the variable $x=E_\gamma/E_p$
Eq.(\ref{Phi g1}) can be written in the following form
\begin{eqnarray}
&\Phi_\gamma(E_\gamma)=&\nonumber\\[5pt]
&\ds c\,n_{H}^{}\int\limits_0^1\! \sigma_{\rm inel}(E_\gamma/x)\,
J_p(E_\gamma/x)\,F_\gamma(x,\,E_\gamma/x)\,\frac{dx}{x}\,.&
\label{Phi g2}
\end{eqnarray}

Analogous equations describe the production of neutrinos $\nu+\tilde\nu$
and electrons $e^-+e^+$; for muonic neutrinos
both contributions from Eqs.(\ref{LEP1}) and (\ref{LEP3})
should be included in calculations.
The inelastic part of the total cross-section of \textit{p-p} interactions
$\sigma_{\rm inel}(E_p)$ can be presented in the following form
\be
\label{sigma inel}
\sigma_{\rm inel}(E_p)=34.3+1.88\,L+0.25\,L^2\,,\; {\rm mb}\,,
\ee
where $L=\ln(E_p/1\,{\rm TeV})$.
This approximation is obtained with the fit of the
numerical data included in the SIBYLL code.

Thus, the calculation of production rates of gamma-rays, electrons
and neutrinos for an arbitrary energy distribution of protons is reduced to
one-dimensional integrals like the one given by Eq.(\ref{Phi g2}).
The analytical presentations described in the previous section can be used,
however, only at high energies: $E_p >0.1\,{\rm TeV}$, and
$x_i=E_i/E_p \geq 10^{-3}$. Continuation of calculations to lower energies
requires a special treatment which is beyond the scope of this paper.
This energy region has been comprehensively studied by Dermer \cite{Dermer86} 
and recently by Kamae et al. \cite{Kamae05}.
On the other hand one may suggest a simple approach which would allow
to continue the calculations, with a reasonable accuracy,
down to the threshold energies of production of particles at
pp interactions. In particular, for distributions of protons 
presented in  the form
\be
\label{pr spectr}
J_p(E_p)=\frac{A}{E_p^\alpha}\,\exp\left[-\left(\frac{E_p}{E_0}\right)^\beta\right]\,,
\ee
the spectra of gamma-rays and other secondaries can be continued to low energies
using the $\d$-functional approximation as proposed  in Ref.\cite{AhAt00}. We suggest a modified version of this approach.
Namely, we adopt for the production rate of $\pi$-mesons
\be
\label{delta1}
\t F_\pi=\t n\,\d\left(E_\pi-\frac{\kappa}{\t n}\,E_{\rm kin}\right),
\ee
where $E_{\rm kin}=E_p-m_p$ is the kinetic energy of protons.
The physical meaning of parameters $\t n$ and $\kappa$
is clear from the following relations:
\be
\int\!\t F_\pi\,dE_\pi=\t n\,,\quad {\rm and}
\quad \int\!E_\pi\,\t F_\pi\,dE_\pi=k\,E_{\rm kin}\, .
\ee
$\t n$ is the number of produced pions for the given
distribution function $\t F_\pi$,  and $\kappa$ is the
fraction of kinetic energy of the proton transferred  
to gamma-rays or leptons. For example for
power-law type distribution functions $\t F_\pi(E) \propto E^{-\alpha}$
with $\alpha \geq 2$, the parameter $\t n \sim 1$.

Assuming that the parameters $\t n$ and $\kappa$ depend weakly
on the proton energy, from Eq.(\ref{delta1}) one finds the production
rate of $\pi^0_{}$-mesons
\be
\label{emiss}
q_\pi(E_\pi)=\t n\,\frac{c\,n_H^{}}{K_\pi}\,\sigma_{\rm inel}\!\left(m_p+\frac{E_\pi}{K_\pi}\right)
J_p\!\left(m_p+\frac{E_\pi}{K_\pi}\right),
\ee
where $K_\pi=\kappa/\t n$. For the procedure described below 
$\t n$ and $K_\pi$  are  free parameters.

The emissivity of gamma-rays is related to
$q_\pi(E_\pi)$ though the equation
\be
\label{emissg}
\frac{dN_\gamma}{dE_\gamma}=2\int\limits_{E_{\min}}^\infty
\frac{q_\pi(E_\pi)}{\sqrt{E_\pi^2-m_\pi^2}}\,dE_\pi\,,
\ee
where $E_{\min}=E_\gamma+m_\pi^2/4E_\gamma$.

The feasibility  of the $\d$-functional approximation in the energy range
$E<100\,{\rm GeV}$ is explained by the following reasons.

1. In the energy range $1\ll E\alt100\,{\rm GeV}$
the cross-section given by Eq.(\ref{sigma inel}) is almost constant,
and the spectrum of protons given by Eq.(\ref{pr spectr})
has a power-law form. Therefore the spectra of gamma-rays and leptons
are also power-law with the same index $\alpha$.
In this case the $\d$-functional approximation leads to power-law spectra
for any choice of parameters $\t n$ and $K_\pi$. Therefore for the given
$K_\pi$ and defining the value of $\t n$ from the condition
of continuity of the spectrum at the point $E=100\,{\rm GeV}$,
one can obtain correct dependence and absolute value of the gamma-ray spectrum
at $1\ll E\alt100\,{\rm GeV}$.

\begin{figure}[h]
\centering{
\includegraphics[width=0.4\textwidth,angle=0,clip=]{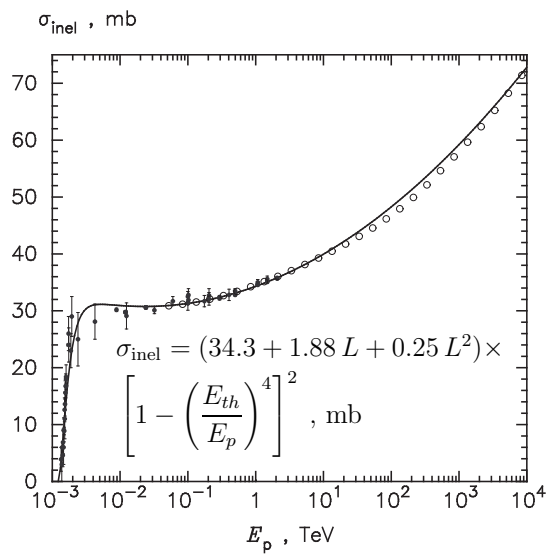}
}
\caption{\small Inelastic cross-secton of \textit{p-p} interactions approximated
by Eq.(\ref{sigma inel1}).  The experimental data are from  
http:wwwppds.ihep.su:8001/c5-5A.HTML,  the open points correspond to 
the cross-sections which are used in the SIBYLL code.}
\label{sigma_in}
\end{figure}

2. For the value of $K_\pi=0.17$, the $\d$-functional approximation for
power-law proton spectra agrees quite well, as is demonstrated  in 
Ref.\cite{AhAt00}, with numerical Monte Carlo calculations \cite{Mori},  
even at energies as low as  $E\sim1\;{\rm GeV}$ (see also discussion in \cite{Torres}).
At lower energies  one has to use, instead of Eq. (\ref{sigma inel}),
a more accurate approximation for the inelastic cross-section:
\begin{eqnarray}
&\sigma_{\rm inel1}(E_p)=(34.3+1.88\,L+0.25\,L^2)\,\times&\nonumber\\[4pt]
&\ds\left[1-\left(\frac{E_{th}}{E_p}\right)^{\!\!4}\right]^2\,,\; {\rm mb}\,,&
\label{sigma inel1}
\end{eqnarray}
where $E_{th}=m_p+2m_\pi+m_\pi^2/2m_p=1.22\cdot10^{-3}\,{\rm TeV}$
is the threshold energy of production of $\pi^0_{}$-mesons.
Eq.(\ref{sigma inel1}) correctly describes the cross-section
also at energies close to the threshold, and at $E_p>3\,E_{th}$
almost coincides with Eq.(\ref{sigma inel}).
The comparison with experimental data \cite{ParticleData}
shows that  Eq.(\ref{sigma inel1})  can be used in  wider energy range of  
protons,   as it is demonstrated  in Fig.~\ref{sigma_in}.

%

\begin{figure*}[t]
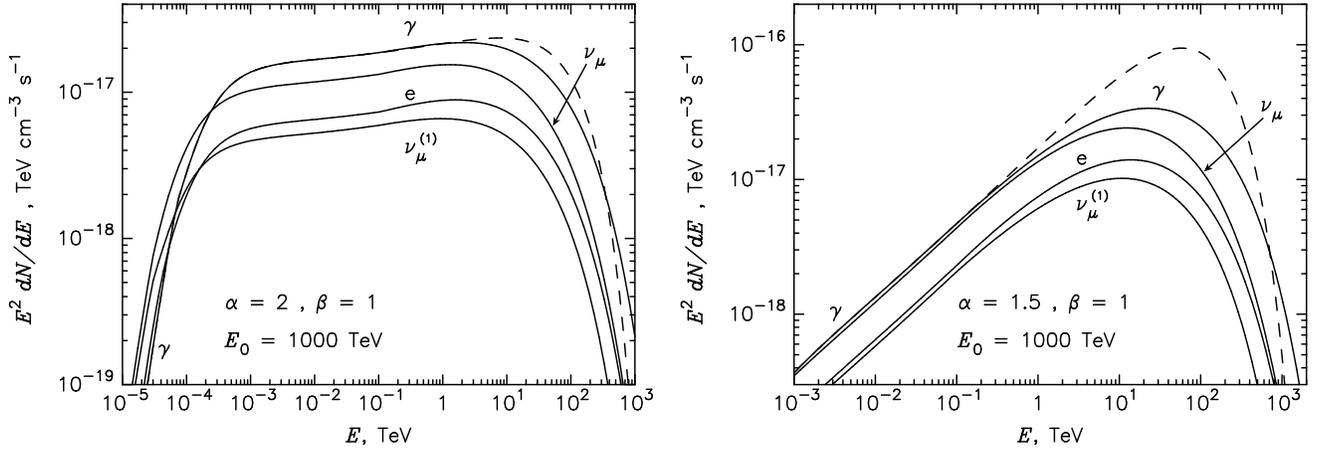

\centering{%
\includegraphics[width=0.33\textwidth,angle=-90,clip=]{12a.eps}\quad
\includegraphics[width=0.33\textwidth,angle=-90,clip=]{12b.eps}
\caption{\small Energy spectra of gamma-rays and leptons from \textit{p-p}
interactions calculated for the distribution of protons given by Eq.(\ref{pr spectr})
with parameters $E_0=$1000~TeV, $\beta=1$ and (a) $\alpha=2$,
(b)$\alpha=1.5$.
The dashed curves are calculated in the $\d$-functional approximation.
}
\label{all_part2}
}
\end{figure*}

In Fig.~\ref{all_part2} we show the spectra of gamma-rays and leptons
calculated for the  proton distribution given by Eq.(\ref{pr spectr}).
The constant $A$ is determined from the condition
\be
\int_{1\,{\rm TeV}}^{\infty}E_p\,J_p(E_p)\,dE_p=1\;{\rm erg}\;{\rm cm}^{-3}\,.
\ee

In the energy range $E\ge 0.1\;{\rm TeV}$
calculations are performed using Eq. (\ref{Phi g2}) with functions
$F_j(x,\,E_p)$  presented  in Section \ref{spectra};
at  lower energies the $\d$-functional approximation
is used with $K_\pi=0.17$. As discussed above,  $\t n$ is treated 
as a free parameter determined from the condition to match the 
spectrum based on accurate calculations at
$E=0.1\,{\rm TeV}$. The parameter $\t n$ depends on the spectrum
of primary protons. For example, for gamma-rays, 
$\t n =1.10$; 0.86, and 0.91 for power-law proton spectra with
$\alpha =2\,;$ 2.5, and 3, respectively. For electrons these numbers are
somewhat different, namely $\t n =0.77$; 0.62, and 0.67 for the
same spectral indices of protons.

In In Fig.~\ref{all_part2} we show also the gamma-ray 
spectrum obtained with the $\d$-functional
approximation in the entire energy range (dashed line).
At low energies where the gamma-ray spectrum behaves like power-law,
the $\d$-functional approximation agrees very well with accurate
calculations. However in the high energy regime where the  proton spectrum deviates
from power-law, the $\d$-functional approximation fails to describe 
correctly  the gamma-ray spectrum.

\begin{figure}[h]
\centering{%
\includegraphics[width=0.3\textwidth,angle=-90,clip=]{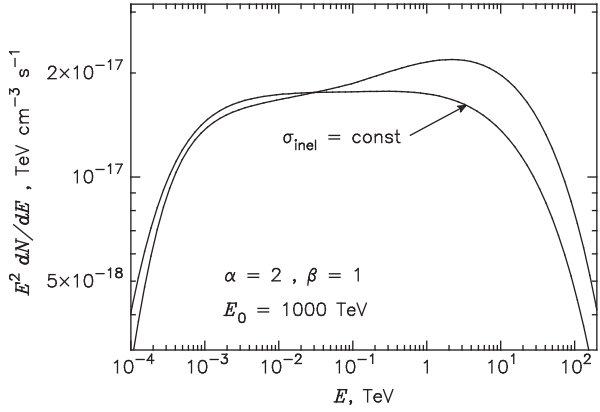}
\caption{\small Comparison of energy spectra of gamma-rays
calculated for the \textit{nominal} inelastic cross-section given by
Eq.(\ref{sigma inel}) and a \textit{formally assumed}
energy independent cross-section $\sigma_{\rm inel}=34\,{\rm mb}$.
The proton distribution is given by Eq.(\ref{pr spectr}), the relevant
parameters are shown in figure.
}
\label{GAM_C}
}
\end{figure}

It should be noted that the noticeable spectral feature (hardening) around
$E\sim 0.1\;{\rm TeV}$ in Fig.~\ref{all_part2}a is not a computational effect
connected with the transition from the $\d$-functional approximation
at low energies to accurate calculations at higher energies.
This behavior is caused, in fact,  by the increase of the
cross-section $\sigma_{\rm inel}$ which becomes
significant at energies of protons above 1 TeV
(see Fig.~\ref{sigma_in}). This is demonstrated in Fig.~\ref{GAM_C}. Indeed, it can be
seen that under a formal assumption of energy-independent cross-section,
$\sigma_{\rm inel}={\rm const}$, the effect of the spectral hardening disappears.

One can seen from Figs.~\ref{all_part2}a,b that the exponential cutoff in the
spectrum of protons at $E_0=10^3$ TeV has an impact on the spectrum of gamma-rays
already at 10 TeV, i.e. as early as $0.01 E_0$.
The reason for such an interesting behavior can be understood from the following
qualitative estimates. The integrand of  Eq.(\ref{Phi g2})
at $E_\gamma\ll E_0$ has sharp maximum in the region of small $x$.
To find the location of the maximum,  let's  assume $\sigma_{\rm inel}={\rm const}$,
and replace the distribution $F_\gamma$ from Eq.(\ref{fit_gamma}) by a model
function $F_\gamma\propto (1-x)^4/x$ which correctly describes the features of
$F_\gamma$ at points $x=0$ and $x=1$.
Then for  $\alpha=2$ and $\beta=1$
the integrand is proportional to the following function
\be
g(x)=(1-x)^4\,e^{-\rho/x}\,,
\ee
where $\rho=E_\gamma/E_0$.
Presenting the function $g(x)$ in the form $g(x)=\exp[h(x)]$,
one can estimate the integral of this function using the saddle-point
integration method,
\be
\int\limits_0^1\!g(x)\,dx\sim\exp[h(x_*)] \ ,
\ee
where $x_*=\frac18\,(\sqrt{\rho^2+16\rho}-\rho)$ is the
location of the maximum of the function $h(x)$. At $\rho\alt1$
one has $h(x_*)\approx -4\,\rho^{1/2}-\rho/2$.
Therefore the spectrum of gamma-rays contains an exponential
term $\exp\left[-(16\,E_\gamma/E_0)^{1/2}\right]$.  The impact of
this term is  noticeable already at energies $E_\gamma\agt10^{-2}\,E_0 $.
This formula also shows that the spectrum of gamma-rays decreases
slower compared to the spectrum of parent protons. In the 
$\d$-functional approximation  the gamma-ray spectrum repeats 
the shape of the proton spectrum. The deviation of the
$\d$-functional approximation from the correct description of the gamma-ray spectrum is clearly seen in Figs.\ref{all_part2}a,b.

\begin{figure*}[t]
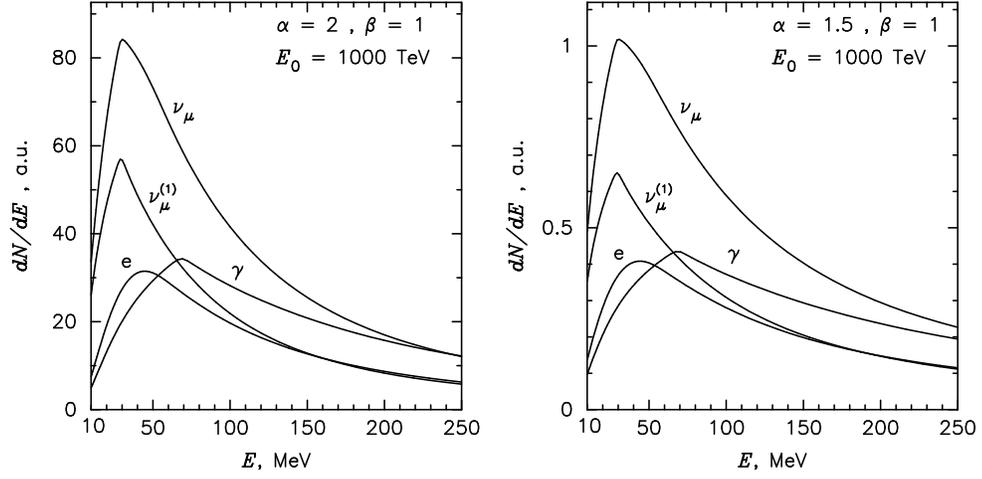

\centering{%
\qquad\includegraphics[width=0.35\textwidth,angle=-90,clip=]{14a.eps}\quad
\includegraphics[width=0.35\textwidth,angle=-90,clip=]{14b.eps}
\caption{\small Energy spectra of gamma-rays and leptons from pp interactions
in the very low energy band. The energy distribution of parent protons is given
by Eq.(\ref{pr spectr}), the relevant parameters $\alpha$, $\beta$, $E_0$
are shown in figures.
}
\label{all_part_ini2}
}
\end{figure*}

Finally we note that for $\alpha=2$ and at energies $E \ll E_0$, the amplitide
of the gamma-ray spectrum exceeds the level of the flux of muonic neutrinos.
The difference is due to the additional
contribution of $\eta$-mesons in production of gamma-rays.
If one takes into account only gamma-rays from decays of $\pi^0$-mesons,
the spectra of gamma-rays and muonic neutrinos are almost identical at energies
well below the cutoff region, where both have power-law behavior. This conclusion
is in good agreement with early studies \cite{Berez91,Gaisser,DAV,Vissani},
and does not contradict to the fact that in each \textit{pp} 
interaction the number of muonic
neutrinos is a factor of 2 larger, on average, than the number of gamma-rays.
The disbalance is compensated, in fact, at low energies.
In Fig.~\ref{all_part_ini2} we show spectra of photons and leptons at
very low energies (see Section \ref{pi-decay}).  It is seen that the number of muonic neutrinos indeed
significantly exceeds the number of gamma-rays. 

%

%

%

\section{Summary}
We present new parameterizations of energy spectra of gamma-rays,
electrons and neutrinos produced in proton-proton collisions.
The results are based on the spectral fits of secondaries produced
in inelastic \textit{p-p} interactions simulated by the SIBYLL code.
The recommended simple analytical presentations provide accuracy better
than 10 percent over the energy range of parent protons $0.1-10^5$ TeV
and $x=E_i/E_p \geq 10^{-3}$.

\begin{acknowledgments}
We are grateful to Venya Berezinsky, Sergey
Ostapchenko,  Francesco Vissani, and especially  Paolo Lipari, 
for very useful comments and
suggestions which helped us to improve significantly the original
version of the paper. We thank Alexander Plyasheshnikov
for his contribution at the early stage of this project and
stimulating discussions.
Finally, we thank the referees for the valuable 
comments and suggestions. S.K and V.B, acknowledge the
support and hospitality of the Max-Planck-Institut f\"ur
Kernphysik during their visit and fruitful collaboration with the
members of the the High Energy Astrophysics group.
\end{acknowledgments}

\appendix
\section{Energy distributions of neutrinos and electrons from   decays of charged pions}

For derivation of the energy distributions of electrons 
and neutrinos at decays of charged pions, 
below we introduce three systems of reference
 -- $K'$, $K_0$, and $K$ 
corresponding to the the muon rest system, the pion rest system and 
the laboratory system, respectively. For certainty, we will consider the $\pi^-$-decay
Since $m_\mu\gg m_e$ we will neglect the mass of the electron. 

Let's denote by $g_e(E'_e,\cos\tilde\theta')\,dE'_e\,d\Omega'$ the energy and 
and angular distribution of electrons in the system $K'$, where 
$\tilde\theta'$ -- is the angle between the muon spin and the electron momentum. 
Since at  the $\pi^-$ decay  the  muon's spin is parallel to its momentum, 
for derivation of the distribution in the   $K_0$ system
one needs  to perform Lorentz transformation towards  the muon's spin. This 
results in 

\[
g^{(0)}_e(E_{e0},\cos\theta_0)\,dE_{e0}\,d\Omega_0=
\frac1{\gamma_\mu(1-\beta_\mu\cos\theta_0)}\times
\]
\be
\label{ap2}
g_e\!\left(E_{e0}\gamma_\mu(1-\beta_\mu\cos\theta_0),\,
\frac{\cos\theta_0-\beta_\mu}{1-\beta_\mu\cos\theta_0}\right)dE_{e0}\,d\Omega_0\,,
\ee
where  $E_{e0}$ is the energy of the electron, $\theta_0$ 
is the angle between the electron and muon momenta, and 
\be\label{ap1}
\gamma_\mu=\frac{1+r}{2\sqrt{r}}\,,\quad \beta_\mu=\frac{1-r}{1+r}\,,
\ee
are the Lorentz factor and speed of the muon - all in the $K_0$ system;  
$r=m_\mu^2/m_\pi^2$.  

In the $K_0$ system the muon can be emitted, with same probability,
at  any angle, therefore  Eq.(\ref{ap2}) should be averaged 
over the muon directions. The function  $g_e$ 
is given by  Eq.(\ref{distr e})  in Sec.\ref{pi-decay}. Then,  denoting 
$\cos\theta_0={\bf n}_e {\bf n}_\mu$ (${\bf n}_e $  and ${\bf n}_\mu$ --
are unite vectors towards the electron and muon momenta), and 
integrating over the directions of the vector ${\bf n}_\mu$,  one obtains
\be\label{pi-dec9}
\bar g_{e0}(E_{e0})\,dE_{e0}\,d\Omega_0=
F_e(\xi)\;d\xi\,\frac{d\Omega_0}{4\pi}\,.
\ee
Here 
\be\label{pi-dec9a}
F_e(\xi)=\Psi_e^{(1)}(\xi)\,\Theta(r-\xi)+\Psi_e^{(2)}(\xi)\,\Theta(\xi-r)\,,
\ee
where $\xi=2E_{e0}/m_\pi$ varies within  $0\le\xi\le1$, and
\be\label{pi-dec10}
\Psi_e^{(1)}(\xi)=\frac{2\xi^2}{3r^2}\,(3+6r-2\xi-4r\xi)\,,
\ee
\be\label{pi-dec11}
\Psi_e^{(2)}(\xi)=\frac{2\,(3-2r-9\xi^2+6r\xi^2+6\xi^3-4r\xi^3)}{3(1-r)^2}\,.
\ee

The spherically symmetric function given by 
Eq.(\ref{pi-dec9})  describes the energy and angular distribution 
of electrons in the pion rest system.   

Similar calculations for electronic neutrinos result in 
\be\label{pi-dec12}
\bar g_{\nu_e0}(E_{e0})\,dE_{\nu_e0}\,d\Omega_0=
F_{\nu_e}(\xi)\;d\xi\,\frac{d\Omega_0}{4\pi}\,.
\ee
In this equation $\xi=2E_{\nu_e}/m_\pi$ ($0\le\xi\le1$), and 
\be\label{pi-dec12a}
F_{\nu_e}(\xi)=\Psi_{\nu_e}^{(1)}(\xi)\,\Theta(r-\xi)+
\Psi_{\nu_e}^{(2)}(\xi)\,\Theta(\xi-r)\,,
\ee
where 
\be\label{pi-dec13}
\Psi_{\nu_e}^{(1)}(\xi)=\frac{4\xi^2}{r^2}\,(3r-\xi-2r\xi)\,,
\ee
\be\label{pi-dec14}
\Psi_{\nu_e}^{(2)}(\xi)=\frac{4\,(-r+3\xi-6\xi^2+3r\xi^2+3\xi^3-2r\xi^3)}{(1-r)^2}\,.
\ee

Now one can calculate, for the distributions of electrons and neutrinos in the 
$K_0$ system  given by Eqs.(\ref{pi-dec9}) and (\ref {pi-dec12}), 
the corresponding energy distributions in the Laboratory system $K$. 
Generally, the exact analytical presentations of these distributions appear quite complex, therefore  in Sec.\ref{pi-decay} we  
present Eqs.(\ref{f1}) -- (\ref{f6}) derived for the 
most interesting case of ultrarelativistic pions.

Finally we should note that the energy distributions of electrons 
published  in  Ref.\cite{Mosk} are not correct. From derivation of Eq.(C2) 
of that paper one can conclude that this equation 
is obtained under the assumption  that the muon and pion 
momenta are strictly parallel, i.e. 
the important step of integration of Eq.(C2) over the muon directions was
erroneously skipped. Consequently,  the integration of Eq.(C2) 
over the energy distribution of muons  leads to inaccurate  Equations (C6) - (C12)
of Ref.\cite{Mosk}.


\end{document}